\documentclass[useAMS,usegraphicx,usenatbib]{mn2e}
 
\usepackage{times}
\usepackage{amssymb}
\usepackage{amsmath}
\usepackage{rotate}
\newif\ifAMStwofonts
\AMStwofontstrue

\newcommand{\Mpc}{\rm\thinspace Mpc}
\newcommand{\kpc}{\rm\thinspace kpc}

\newcommand{\km}{\rm\thinspace km}

\newcommand{\keV}{\rm\thinspace keV}

\newcommand{\kmps}{\hbox{$\km\s^{-1}\,$}}

\newcommand{\kmpspMpc}{\hbox{$\kmps\Mpc^{-1}$}}

\newcommand{\ps}{\hbox{$\s^{-1}\,$}}

\newcommand*{\prog}[1]{\textsc{#1}}
\newcommand*{\code}[1]{\texttt{#1}}
\newcommand*{\file}[1]{\textsc{#1}}

\newcommand*{\projct}{\code{projct}}
\newcommand*{\phabs}{\code{phabs}}
\newcommand{\xmekal}{\code{mekal}}

\newcommand*{\dmextract}{\prog{dmextract}}

\newcommand*{\lcclean}{\prog{lc\_clean}}
\newcommand*{\mkwarf}{\prog{mkwarf}}
\newcommand*{\mkrmf}{\prog{mkrmf}}

\newcommand*{\pha}{\file{pha}}

\newcommand*{\xspec}{\prog{xspec}}

\def\mrk766{{Mrk 766}}

\def\ps{{\rm\thinspace s^{-1}}}
\def\km{{\rm\thinspace km}}

\def\kmps{\hbox{$\km\ps\,$}}
\def\Mpc{{\rm\thinspace Mpc}}
\def\kmpspMpc{\hbox{$\kmps\Mpc^{-1}\,$}}

\title{Galaxy cluster mass profiles} \author[Voigt \& Fabian]
{\parbox[]{6.in} {L.M. Voigt and A.C. Fabian\\
    \footnotesize
    Institute of Astronomy, Madingley Road, Cambridge CB3 0HA\\
  }}

\begin{document}
 
\maketitle
 
\label{firstpage}

\begin{abstract} Accurate measurements of the mass distribution in
  galaxy and cluster halos are essential to test the cold dark matter
  (CDM) paradigm. The cosmological model predicts a universal shape
  for the density profile in all halos, independent of halo mass.  Its
  profile has a `cuspy' centre, with no evidence for the constant
  density core. In this paper we carry out a careful analysis of
  twelve galaxy clusters, using \emph{Chandra} data to compute the
  mass distribution in each system under the assumption of hydrostatic
  equilibrium. Due to their low concentration, clusters provide ideal
  objects for studying the central cusps in dark matter halos. The
  majority of the systems are consistent with the CDM model, but 4
  objects exhibit flat inner density profiles. We suggest that the
  flat inner profile found for these clusters is due to an
  underestimation of the mass in the cluster centre (rather than any
  problem with the CDM model), since these objects also have a
  centrally peaked gas mass fraction. We discuss possible causes for
  erroneously low mass measurements in the cores of some systems.
  \end{abstract}
 
\begin{keywords}
  galaxies: clusters -- X-rays: galaxies -- dark matter
\end{keywords}

\section{Introduction}

In hierarchical collapse models, $N$-body simulations predict a
universal shape for the mass distribution in dark matter halos from
dwarf galaxies (10$^{7}$ M$_{\odot}$) to massive clusters (10$^{15}$
M$_{\odot}$), independent of the value of the cosmological parameters.
The density profile differs strongly from a simple single power-law,
predicted in early theoretical work
\citep{gunn&gott1972,fillmore&goldreich1984,hoffman&shaham1985,white&zaritsky1992},
and is centrally concentrated with an inner cusp \(\rho(r) \propto
r^{-\alpha}\). The universality of the structure of dark matter halos
formed through hierarchical clustering was `discovered' in simulations
performed by \citet{nfw1995,nfw1996,nfw1997}, hereafter
NFW\footnote{Simulations by NFW built on earlier, pioneering works by
  \citet{frenketal1988}, \citet{dubinski&carlberg1991} and
  \citet{croneetal1994}, which identified the absence of a constant
  density core in dark matter halos.}. In these early studies, an
inner slope with $\alpha=1.0$ provided the best-fit to the spherically
averaged density profiles.  This was supported by various authors
\citep{cole&lacey1996,tormenetal1997,kravtsovetal1997}.  Later
simulations at higher resolution indicated a steeper central cusp,
with $\alpha=1.5$
\citep{fukushige&makino1997,fukushige&makino2001,fukushige&makino2003,mooreetal1999,ghignaetal2000,klypinetal2001}.
A density profile similar to the NFW profile, but with $\alpha=1.5$
was suggested by \citet{mooreetal1999}, hereafter M99.

A thorough understanding of the predictions made by simulations is
essential in order to test the cold dark matter (CDM) paradigm against
observations. There are several issues which need to be addressed by
simulators, in particular in the high density central regions of halos
where large numbers of particles and fine time resolution are
required. In recent work there has been some debate over the form of
the profile at the innermost resolved radii ($r < 0.01r_{\rm vir}$).
\citet{poweretal2003}, \citet{hayashietal2003} and
\citet{navarroetal2004} suggest that the inner density profile does
not converge to a well-defined power-law, but continues to flatten
inwards. \citet{diemandetal2005}, on the other hand, dispute this
result, continuing to support the existence of a central asymptotic
power-law to the density profiles.

A further issue which remains unresolved numerically is the
universality of the core profile.  \citet{jing&suto2000}, for example,
find that the inner slope steepens with decreasing halo mass.  In
addition, using analytical arguments, \citet{hoffman&shaham1985} and
\citet{syer&white1998} show that the halo profile should depend on the
power spectrum of initial density fluctuations, with an inner slope
given by \(3(n+3)/(n+4)\) and \(3(n+3)/(n+5)\), respectively, where
$n$ is the power-law index of the spectrum.
 
Nevertheless, there are two robust predictions made by simulations
which may be tested against observations: the density profile in CDM
halos differs strongly from a single power-law, and the inner region
is cuspy with a power-law slope in the range $1.0 \lesssim \alpha
\lesssim 1.5$ down to at least 1 per cent of the virial radius.  In
this paper we obtain mass profiles under the assumptions of
hydrostatic equilibrium and spherical symmetry for a sample of X-ray
peaked galaxy clusters using \emph{Chandra} data. Straightforward
deprojection techniques are robust and work well on such clusters.  In
addition, clusters halos have smaller concentrations than galaxies,
making them ideal objects to study cusps.  Work on the inner density
slope of clusters of galaxies has also been carried out in recent
studies by \citet{katayama&hayashida2004}, \citet{aba2004} and
\citet{pap2005}.

A $\Lambda$CDM cosmology with $\Omega_{\rm \Lambda} = 0.7$ and $h =
0.7$ is adopted.

\section{Cluster sample} 

The cluster sample used is listed in Table~\ref{clusters}.
Observations of objects with known short central cooling times,
indicating a relatively relaxed object, were obtained from the
\emph{Chandra} archive\footnote{Observations using the ACIS-S3
  detector only were used.}. The sample of 12 objects was chosen to
cover a range in cluster redshift ($z=0.02-0.46$) and ICM temperature
($k_{\rm B}T=2-15 \keV$).  High redshift clusters observed by
\emph{Chandra} tend to be hotter, and in this sense the sample is not
statistically complete; however, this will not in general affect the
analyses carried out, and any biases introduced by the limited sample
are discussed where relevant.

Data processing was carried out using the CIAO (\emph{Chandra
  Interactive Analysis of Observations}) software package available
from the CXC (\emph{Chandra X-ray Centre}).  Level 2 events files were
updated to include the latest calibration.  Only those X-ray events
with \emph{ASCA} grade classifications 0, 2, 3, 4 and 6 were included
in the cleaned data set. Further screening was carried out to remove
time intervals contaminated by background flares.  The lightcurves
were analysed using the script \lcclean{} written by M.~Markevitch and
provided by the CXC. The observation length remaining after screening
(GTI; good time interval) is listed for each object in
Table~\ref{clusters}.

\begin{table*}
  \begin{center}
\begin{tabular}{l c c r@{.}l r@{.}l c c r@{.}l c}
\hline
\\
Cluster& Redshift & Obs. date& \multicolumn{2}{c}{Exps. time} & \multicolumn{2}{c}{GTI}&
\multicolumn{2}{c}{Emission peak (J2000)} &  \multicolumn{2}{c}{$D_{\rm L}$} &  $D_{\rm A}$ \\
&  & &\multicolumn{2}{c}{(ks)} &\multicolumn{2}{c}{(ks)}
&RA & Dec  & \multicolumn{2}{c}{(Mpc)} &(kpc arcsec$^{-1}$) \\
\hline
Abell~3112& 0.0750 & 2001 Sep 15 & ~~~~~17&1 & 13&5 & 03 17 57.7 & $-$44 14
16.9 & 335&2 & 1.4 \\
2A 0335$+$096 & 0.0347 & 2000 Sep 06 & 20&0 & 18&1 & 03 38 40.6 &  $+$09 58
11.0 &140&6 & 0.7\\
Abell~478 & 0.0880  & 2001 Jan 27 &  42&9 &38&9 & 04 13 25.2 &  $+$10 27 53.9
&396&9 &1.6  \\
PKS~0745$-$191 & 0.1028 & 2001 Jun 16 & 17&9 &14&6 &07 47 31.2 &  $-$19 17
38.8 &468&5 & 1.9 \\
RXJ~1347.5$-$1145(1)& 0.4510& 2000 Mar 5&  9&1 & 8&0& 13 47 30.5 &  $-$11 45
09.5 &2492&9 &5.7 \\
RXJ~1347.5$-$1145(2)& 0.4510 &2000 Apr 29 & 10&1 & 7&2 & 13 47 30.5  & $-$11 45
09.5 & 2492&9 & 5.7  \\
Abell~1795 & 0.0632 &2000 Mar 21  &19&7 &15&6 &13 48 52.5 &  $+$26 35 37.8
&280&0 &1.2 \\
Abell~1835 & 0.2523  & 1999 Dec 11 &19&8 &18&9 & 14 01 01.9  &  $+$02 52 43.4
&1262&4 &3.9 \\
Abell~3581 & 0.0218 & 2001 Jun 07 &7&3 & 6&0& 14 07 29.8 & $-$27 01
04.2 &93&6 & 0.4 \\
Abell~2029 & 0.0767 & 2000 Apr 12 &19&9 & 19&8 &15 10 56.1 &  $+$05 44 40.6
&343&2 &1.1 \\
RXJ~1532.9$+$3021 & 0.3615 & 2001 Aug 26 &9&5 & 6&2 &15 32 53.8  & $+$30
20 58.5 &1916&3 & 5.0 \\
MS~2137.3$-$2353 & 0.3130 & 2000 Dec 10 & 44&2 & 19&5 & 21 40 15.2 &
$-$23 39 39.9 & 1618&8 & 4.7 \\
Sersic~159$-$03 (AS 1101)& 0.0564  &2001 Aug 13 &10&1 &9&8 &23 13 58.3
 & $-$42 43 35.0 &248&7 &1.1 \\
\hline
\end{tabular}
\caption{Summary of the \emph{Chandra} observations. Cluster redshift,
  observation date, exposure time, good time interval, X-ray emission
  peak, luminosity distance and angular scale. Both observations
  available for RXJ~1347.5$-$1145 were used.}
\label{clusters}
\end{center}
\end{table*}

\section{Spectral analysis}

\subsection{Extracting spectra}

For the spectral analysis, each cluster was divided into circular
annuli around the X-ray emission peak, with strong point sources
identified by eye and masked out. Regions disturbed by, for example,
merging subclusters and radio lobes, were also removed (see Section
3.4). A spectrum was extracted for each annulus using the CIAO
\dmextract{} tool.  The spectra were binned to contain at least 20
counts per \pha{} channel to enable the use of chi-square statistics.
Ancillary-response and response matrices were constructed using the
CIAO \mkwarf{} and \mkrmf{} programs.

\subsection{The spectral model}

Spectra were fit in the $0.5-7.0 \keV$ energy range using the \xspec{}
\citep{arnaud96} software package. The emission from each shell was
modelled using the \xmekal{} \citep{mew&gron&van1985} plasma emission
code, incorporating the Fe L calculations of
\citet{lied&ost&gold1995}, and absorbed by the \phabs{}
\citep{bal&mc1992} photoelectric absorption code to take into account
galactic absorption along the line-of-sight.  The emission was
deprojected using the \projct{} model provided in \xspec{}.

The free parameters in the model were the temperature, metallicity and
emission measure. The elements were assumed to be present in the solar
ratios measured by \citet{anders&grev1989} and the abundance allowed
to vary between shells.  The Galactic absorption column density was
left as a free parameter in the fits\footnote{The low energy quantum
  efficiency of the ACIS chips has been continuously degraded since
  launch. Tools have been provided by the CXC to account for the loss
  in effective area, although whether or not the correction is applied
  has a negligible effect on the temperature and emission integral
  profiles when the galactic absorption column density is left as a
  free parameter in the fits. With the correction applied, the
  best-fitting $N_{\rm H}$ is often much less than the nominal value,
  suggesting that the tools tend to `over-correct' the data. Following
  the work of \citet{voigt&fabian2004} and \citet{birzanetal2004}, we
  use uncorrected data in order to avoid best-fitting galactic
  absorption measurements consistent with zero.}, although linked
between shells.  The nominal \citep[measured by][]{dick&lock1990} and
best-fitting $N_{\rm H}$ values are shown in Table~\ref{tab:spectra},
together with the minimum chi-square of the fit.

\begin{table*}
\centering
\begin{tabular}{lcccc}
\hline
\\
Cluster &  Fitted $N_{\rm H}$  & Galactic $N_{\rm H}$ &
$\chi^{2}$(dof) & $\chi^{2}_{\nu}$\\
&  (10$^{20}$ cm$^{-2}$) & (10$^{20}$ cm$^{-2}$) & \\
\hline
Abell~3112& 5.42$^{+0.22}_{-0.22}$ & 1.95 & 1482(1388) & 1.07\\
2A~0335$+$096 & 27.23$^{+0.18}_{-0.18}$ &  17.8 & 2682(1968) & 1.36 \\
Abell~478 & 33.46$^{+0.16}_{-0.16}$ & 15.2 &  4243(3458)  & 1.23\\
PKS~0745$-$191 & 42.77$^{+0.35}_{-0.34}$  &42.4 & 2486(2251) & 1.10 \\
RXJ~1347.5$-$1145 & 5.18$^{+0.72}_{-0.41}$ &4.85 & 805(859) & 0.94 \\
Abell~1795 & 1.82$^{+0.13}_{-0.13}$ & 1.19 & 2110(1861) & 1.13  \\
Abell~1835 & 2.78$^{+0.24}_{-0.24}$ & 2.32 & 1551(1445) &1.07 \\
Abell~3581 &  9.87$^{+0.53}_{-0.50}$  &4.52 & 716(653) &1.10  \\
Abell~2029 & 4.14$^{+0.12}_{-0.11}$ &3.05 & 2878(2555) & 1.13\\
RXJ~1532.9$+$3021 & 6.14$^{+0.68}_{-0.65}$  &2.16 & 403(379) &1.06 \\
MS~2137.3$-$2353 & 4.63$^{+0.44}_{-0.45}$ & 3.55& 596(541) & 1.10 \\
Sersic~159$-$03 & 6.07$^{+0.39}_{-0.39}$& 1.79 & 917(844) &1.09  \\
\hline
\end{tabular}
\caption{Best-fitting absorption column density, $N_{\rm H}$,
  chi-square and reduced chi-square for the model
  \projct*\phabs(\xmekal{}).
  The nominal galactic absorption column along the line-of-sight to
  the cluster from \citet{dick&lock1990} is also tabulated.}
\label{tab:spectra}
\end{table*}

\subsection{Deprojection}

A series of tests have been carried out by \citet{johnstoneetal2004}
to check that \projct{} produces the correct results when applied to
known synthetic data. With the exception of the outermost shell, the
temperature and density profiles obtained with \projct{} match very well
with the synthetic input profiles.

An apparent increase in density in the outermost shell results from
the assumption of zero emission exterior to that region. Clearly,
there will be some counts attributed to this shell which were emitted
at larger radii and the emission integral will be overestimated. This
effect will be seen out to large radii --- even if the cluster counts
exterior to the outer region are negligible, the background emission
will contribute to the emission integral --- although will be become
less significant as the virial radius is approached.

We note that the measured gas temperature in the penultimate annulus
is slightly higher when the outer shell is removed from the fit,
although consistent within the one sigma limits. The temperature in
the outer annulus will be slightly overestimated if the temperature
rises exterior to this region and underestimated if it falls.

\subsection{Masked regions}

To obtain an accurate measure of the cluster mass distribution we are
interested in the ambient gas properties.  For the spectral analyses,
the following features were therefore masked out from the images: a
shock front in 2A~0335$+$096 spanning a sector between
$150-210^{\circ}$ $50-60$ arcsec from the cluster centre; radio lobes
in Abell~478; the southeast quadrant in RXJ~1347.5$-$1145 containing a
bright subclump; the bright filament extending southwards from the
core in Abell~1795.

\subsection{Spectral fits}
The reduced chi-square of the fits obtained using the model
\projct*\phabs(\xmekal{}) are less than $\sim$1.1 for the majority of
objects in the sample (see Table~\ref{tab:spectra}), showing that the
X-ray emission is well described by a single phase plasma at each
radius. We note that the reduced chi-square is greatest for the fits
to 2A~0335$+$096 and Abell~478.

\section{Temperature, density and pressure profiles}

\begin{figure*}
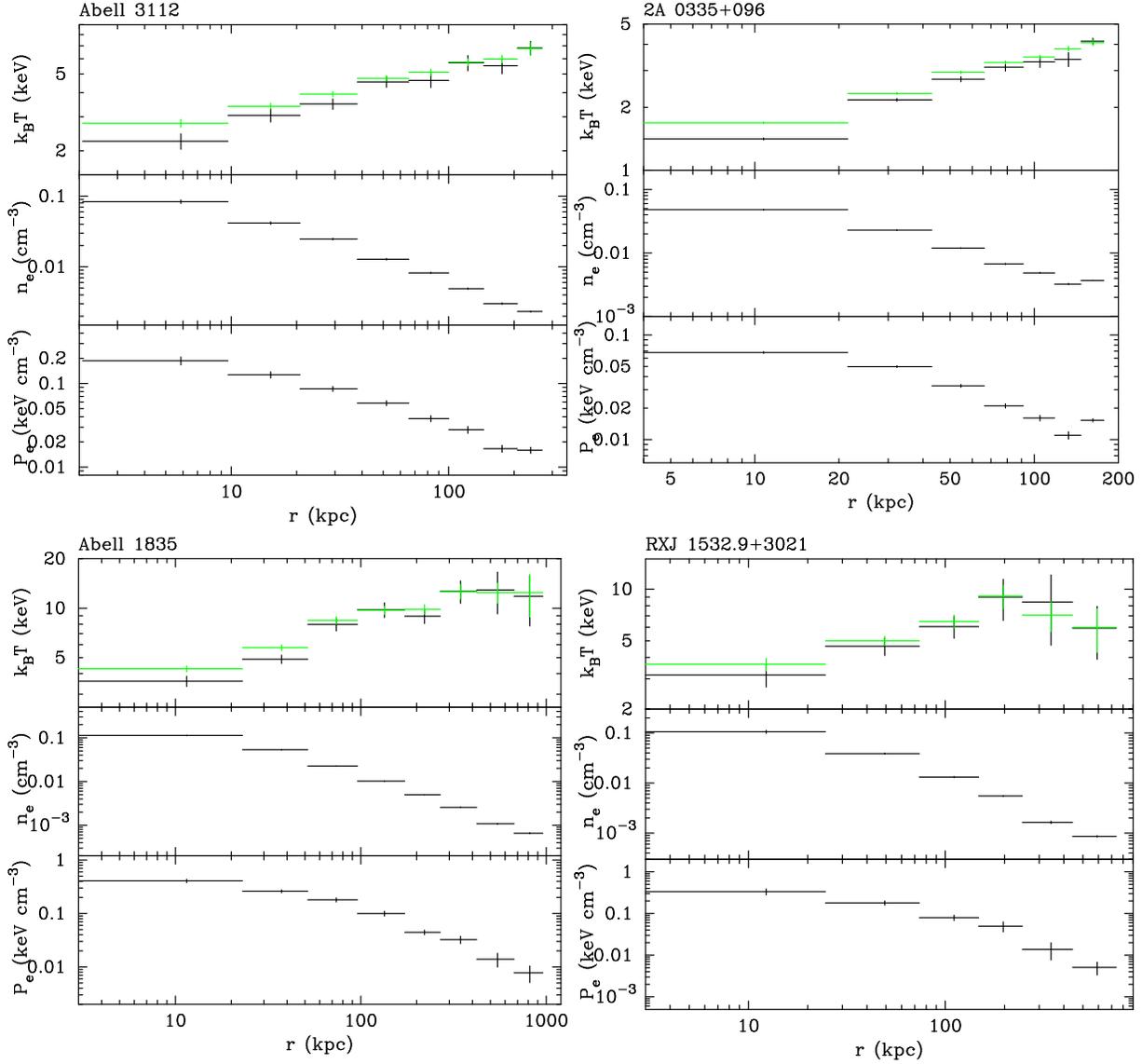

\includegraphics[angle=-90,width=0.95\columnwidth]{a3112temp_ne_prod.ps}
\vspace{1.0mm}
\includegraphics[angle=-90,width=0.95\columnwidth]{2a0335temp_ne_prod.ps}
\vspace{1.0mm}
\includegraphics[angle=-90,width=0.95\columnwidth]{a1835temp_ne_prod.ps}
\vspace{1.0mm}
\includegraphics[angle=-90,width=0.95\columnwidth]{rxj1532temp_ne_prod.ps}
\caption{Deprojected gas temperature, density and pressure profiles
  for a representative sample of clusters. Projected gas temperature
  profiles are shown in green.}
\label{fig:profiles}
\end{figure*}

Gas temperature, density and pressure profiles are shown for a
representative sample of clusters in Fig~\ref{fig:profiles}.

\subsection{Temperature and density}
The temperature distribution is obtained directly from the spectral
fits and the density profile from the normalization of the \xmekal{}
spectrum, \(K=EI/(4\times10^{14}\pi D_{\rm A}^{2} (1+z)^{2})\), where
$EI (= \int n_{\rm e} n_{\rm H} \mathrm{d} V)$ is the emission
integral. The electron number density at the centre of each shell is
estimated as \(\left<n_{\rm e}\right> = (1.2EI/V)^{1/2}\), where we
have used the relation $n_{\rm e} \approx 1.2n_{\rm H}$ and $V$ is the
volume of the shell.

\subsection{Pressure}
At the high temperatures and low densities in the ICM the appropriate
pressure equation of state is the ideal gas law, given by \(P = n
k_{\rm B}T\), where $n \approx 1.92n_{\rm e}$ is the number of
particles per unit volume. The electron pressure, $P_{\rm e}$, at the
midpoint of each shell is computed by multiplying the electron density
by the temperature. The uncertainties in $n_{\rm e}$ and $T$ are added
in quadrature to find the uncertainty in $P_{\rm e}$ (i.e. the density
and temperature are assumed to be independent. We note that contour
plots of shell normalization against temperature are approximately
circular, showing the assumption is a reasonable one).

\section{Mass measurements}

\subsection{Total mass}

Determining masses using X-ray data allows the cluster mass to be
computed as a function of radius. The gas is assumed to be in
hydrostatic equilibrium within the cluster potential.  The equation of
hydrostatic equilibrium for a spherically symmetric system is given by
 
\begin{equation}
M_{\rm tot}(<r_{j})=-\frac{1}{G}\frac{r_{j}^{2}}{\rho_{{\rm gas}, j}} \left(\frac{dP}{dr}\right)_{j},
\label{eqn:hydrostat}
\end{equation}
where $P$ is the thermal pressure.

\subsubsection{Calculation}
The total mass enclosed within radius $r_{j}$ is calculated from
Equation~\ref{eqn:hydrostat}, where $r_{j}$ is the radial distance
from the centre of the cluster to the midpoint between two consecutive
shells. The pressure gradient at the midpoint is estimated as
 
\begin{equation}
\left(\frac{dP}{dr}\right)_{j} = \frac{P_{i+1} - P_{i}}{r_{i+1} -
  r_{i}},
\label{eqn:pressuregrad}
\end{equation}
where $P_{i}$ is the pressure at the centre of the $i$th shell,
$r_{i}$ is the radial distance to the centre of the $i$th shell and
$r_{j} = (r_{i} + r_{i+1})/2$. The gas density at $r_{j}$ is
calculated using linear interpolation, such that
 
\begin{equation}
\rho_{\rm gas}(r_{j}) = \frac{\rho_{\rm gas}(r_{i+1})+\rho_{\rm gas}(r_{i})}{2},
\end{equation}
where $\rho_{{\rm gas}, i}$, the gas density at the centre of the
$i$th shell, is found from the electron number density using the
relation $\rho_{{\rm gas}, i} = \mu m_{\rm H} n_{{\rm gas}, i} \approx
1.92 \mu m_{\rm H} n_{{\rm e}, i}$. The uncertainties are calculated
using a Monte Carlo technique whereby both the pressure and density
data are perturbed 1000 times.  PKS~0745$-$191 does not exhibit a
smooth pressure profile in the centre and so measurements made using
Equation~\ref{eqn:pressuregrad} do not produce a monotonically
increasing mass profile. We compute the mass profile in the inner
region of this object by binning together the first and second and the
third and fourth data points in the pressure and gas density profiles.

Computing the mass profile involves determining the pressure gradient
at the midpoint between the centres of consecutive shells. By
estimating the gradient using Equation~\ref{eqn:pressuregrad} we
effectively fit a straight line between adjacent data points. (The
accuracy of the method may therefore be improved by increasing the
number of data points).  Neighbouring mass points may be slightly
correlated, but data points further away than one will not be
correlated with one another.  Several authors fit a parametric model
to the temperature and gas density profiles and use this to compute
the mass profile. The data points in this case are not independent and
finding the best-fit model to the mass profile using chi-square
statistics will be statistically invalid. The method used here, made
possible due to the high spatial resolution of the \emph{Chandra}
observatory, therefore provides an improvement over previous analyses.
 
\begin{figure*}
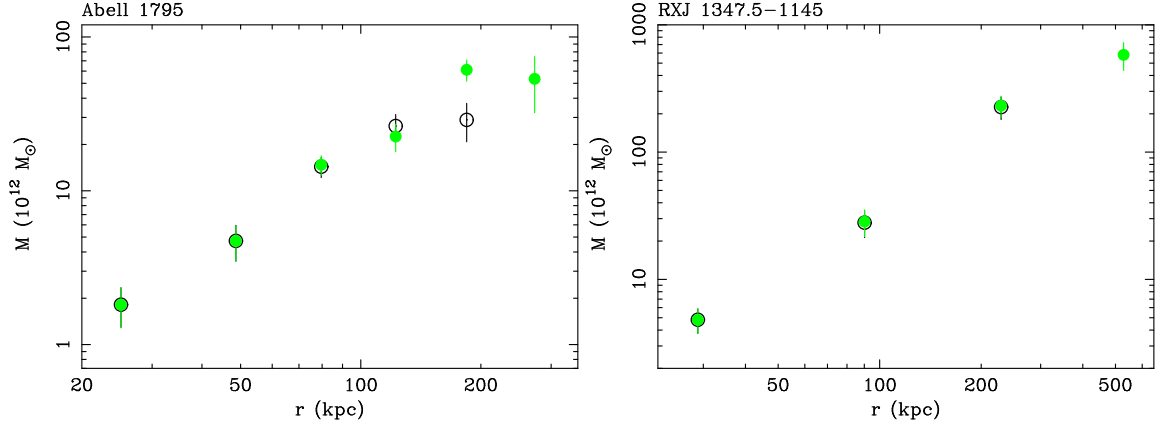

\includegraphics[angle=-90,width=0.9\columnwidth]{a1795massrms_data_1e12_6_7shells.ps}
\vspace{1.0mm}
\includegraphics[angle=-90,width=0.9\columnwidth]{rxj1347massrms_data_1e12_5_4shells.ps}
\caption{Mass profiles computed using data from spectral fits obtained
  with (green filled circles) and without (black open circles) the
  outer shell included.}
\label{fig:massdiff}
\end{figure*}

The mass profiles are shown for the sample in Fig.~\ref{fig:mass}. The
outer mass data point is removed in all but three clusters
(RXJ~1347.5$-$1145, RXJ~1532.9$+$3021 and MS~2137.3$-$2353).  This is
because the artificially high gas density and pressure in the outer
bin\footnote{See Section 3.3.} causes the outer mass data point to be
significantly underestimated in the majority of objects. For the high
redshift systems ($z \gtrsim 0.3$) the outer mass data point is
estimated to be accurate to within a few per cent. The accuracy of the
outer data point is assessed by calculating the mass profile from the
data obtained both with and without the outer shell included in the
spectral analysis.  As an example, the mass profiles obtained for
Abell~1795 and RXJ~1347.5$-$1145 are shown in Fig.~\ref{fig:massdiff}.
The percentage decrease in the mass measured in the penultimate shell
is less than 1 per cent for RXJ~1347.5$-$1145 and about 50 per cent
for Abell~1795 when the outer annulus is not included in the spectral
fit.
 
Removing the outer mass data point in clusters such as Abell~1795
before analysing the mass profile is vital. Current structure
formation models predict that the mass profile should `turn-over' at
some characteristic radius. If an erroneously low mass data point is
included in the outer region then this will wrongly indicate that the
profile is curving away from a power-law.
 
\subsubsection{Lensing masses}

Mass measurements from strong lensing analyses are also plotted in
Fig.~\ref{fig:mass} for PKS~0745$-$191, RXJ~1347.4$-$1145, Abell~1835
and MS~2137.3$-$2353 \citep{allen1998}. In each case the lensing mass
is larger than the X-ray mass by a factor of between ~1.5$-$3. This
discrepancy is also found by \citet{allen1998} if a single phase
spectral model (i.e. no cooling flow) is adopted. It is difficult at
this stage to draw any strong conclusions about the difference in the
measurements. The lensing results are quoted without uncertainties and
there are several difficulties with this method, such as adopting the
correct geometry, obtaining accurate arc redshifts and the unknown
presence of secondary matter along the line of sight
\citep{wambetal2005}.  Similarly, the masses obtained from the X-ray
analysis are also dependent on several assumptions, including the
spectral and geometrical models adopted. It will be important in
future work to improve on both methods until an agreement is reached.

\subsection{Gas mass and gas mass fraction}

The integrated gas mass profile is calculated using the sum

\begin{equation}
  M_{\rm gas}(<r_{k})=\sum_{i=1}^{n} \frac{4}{3} \pi
  (r_{k}^{3}-r_{k-1}^{3}) \rho_{\rm gas}(r_{i}),
\label{eqn:gas}
\end{equation}
where $r_{k}$ is the radial distance to the outer boundary of the
$i$th shell.

\begin{figure*}
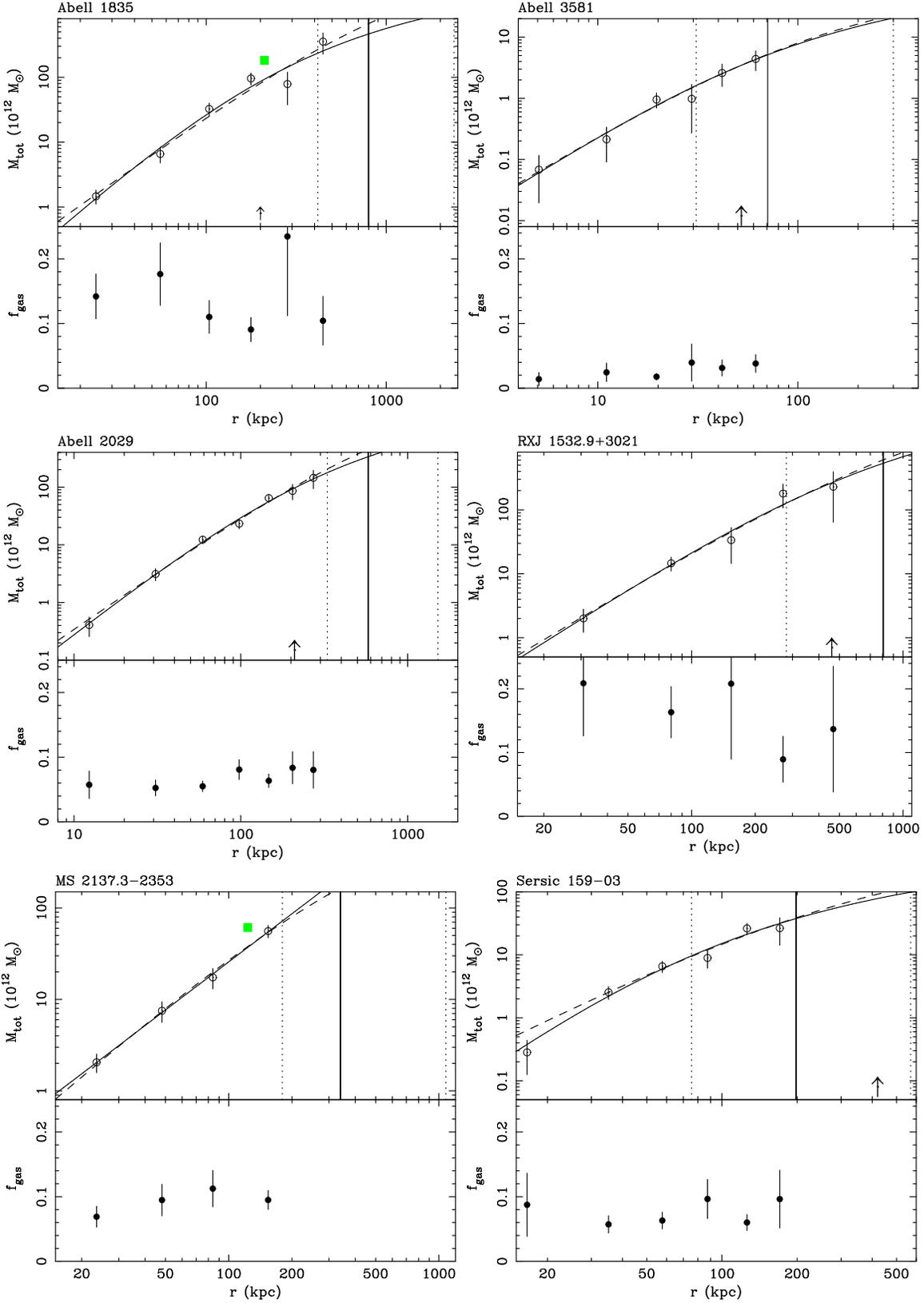

\includegraphics[angle=-90,width=0.9\columnwidth]{a1835massplot.ps}
\vspace{1.0mm}
\includegraphics[angle=-90,width=0.9\columnwidth]{pks1404massplot.ps}
\vspace{1.0mm}
\includegraphics[angle=-90,width=0.9\columnwidth]{a2029massplot.ps}
\vspace{1.0mm}
\includegraphics[angle=-90,width=0.9\columnwidth]{rxj1532massplot.ps}
\vspace{1.0mm}
\includegraphics[angle=-90,width=0.9\columnwidth]{ms2137massplot.ps}
\vspace{1.0mm}
\includegraphics[angle=-90,width=0.9\columnwidth]{ser159massplot.ps}
\vspace{1.0mm}
\caption{Total (dark plus baryonic) mass profiles (upper) and gas mass
  fraction profiles (lower). The error bars show the 1 sigma
  uncertainties in the measurements. The solid line shows the best-fit
  M${_{2}F_{1}}$ model to the data, with $\alpha$ a free parameter in
  the fit. The dashed line shows the best-fit NFW model ($\alpha=1$)
  to the data.  For the latter model the inner 30 kpc is removed from
  the fit to the profiles of 2A~0335$+$096, Abell~478, PKS~0745$-$191
  and Sersic~159$-$03 (see text). The best-fit scale radius and 1
  sigma limits, obtained with $\alpha = 1$, are shown by solid and
  dotted vertical lines respectively. The arrow indicates the best-fit
  scale radius for the M${_{2}F_{1}}$ model. (In the cases where an
  arrow is not shown the best-fit model tends to a power-law). Lensing
  masses taken from \citet{allen1998} are indicated (green filled
  squares).}
\label{fig:mass}
\end{figure*}

The gas mass fraction is the ratio of the total gas mass to the total
gravitating mass within a fixed volume

\begin{equation}
f_{\rm gas} = \frac{M_{\rm gas}(<r_{j})}{M_{\rm tot}(<r_{j})}
\end{equation}
The total gas mass within $r_{j}$ is computed from the gas mass
profiles using linear interpolation. The gas mass fraction profiles
are plotted in Fig.~\ref{fig:mass}.

We note that the total integrated mass within a particular volume,
calculated from the equation of hydrostatic equilibrium, is dependent
upon the pressure gradient and gas density at that radius only, and is
completely unaffected by the regions interior (or exterior) to that
radius. For the gas mass profile, on the other hand, the integrated
mass is calculated by summing from the centre outwards and any error
in the measurement at small radii will propagate out to larger radii.
However, the gas mass at small radii is much less than at large radii
and any uncertainty in the measurements in the core are unlikely to
have a significant effect on the gas mass profile further out.

\begin{figure*}
\includegraphics[angle=-90,width=0.9\columnwidth]{a3112massplot.ps}
\vspace{1.0mm}
\includegraphics[angle=-90,width=0.9\columnwidth]{2a0335massplot.ps}
\vspace{1.0mm}
\includegraphics[angle=-90,width=0.9\columnwidth]{a478massplot.ps}
\vspace{1.0mm}
\includegraphics[angle=-90,width=0.9\columnwidth]{pks0745massplot.ps}
\vspace{1.0mm}
\includegraphics[angle=-90,width=0.9\columnwidth]{rxj1347massplot.ps}
\vspace{1.0mm}
\includegraphics[angle=-90,width=0.9\columnwidth]{a1795massplot.ps}
\end{figure*}

\section{Dark matter profiles from simulations}
 
The generalized equation for the density distribution in dark matter
halos, including all models with a `cuspy' core, is given by
 
\begin{equation}
\rho(x)=\frac{\rho_{0}}{x^{\alpha}\left(1+x^{\gamma}\right)^{(\beta-\alpha)/\gamma}},
\label{eqn:darkden}
\end{equation}
where \(x=r/r_{\rm s}\). The scale radius, $r_{\rm s}$, is a free
parameter. The indices ($\alpha$,$\beta$,$\gamma$) for the NFW and M99
models are (1.0,3.0,1.0) and (1.5,3.0,1.5), respectively. A simple
analytic model for the mass is easily obtained for each of these
special cases, with

\begin{equation}
M(< x)=
\begin{cases}
4 \pi \rho_{0} r_{\rm
  s}^{3}\left[\mathrm{ln}(1+x)-\frac{x}{1+x}\right]& \text{for NFW}\\
\frac{8}{3} \pi \rho_{0} r_{\rm
  s}^{3}\mathrm{ln}(1+x^{3/2}) & \text{for M99}
\end{cases}
\label{eqn:darkmass}
\end{equation}

A generalized mass profile, valid for any values of the indices, may
be written
 
\begin{equation}
M(< x) = 4 \pi \rho_{0} r_{\rm
  s}^{3} \frac{x^{\alpha_{3}}{_{2}F_{1}}(\alpha_{3}/\gamma,(\beta-\alpha)/\gamma;(\alpha_{3}+\gamma)/\gamma;-x^{\gamma})}{\alpha_{3}},
\label{eqn:darkgen}
\end{equation}
where $\alpha_{3}=3-\alpha$. The hypergeometric function ${_{2}F_{1}}$
may be evaluated numerically. We refer to this model as
M${_{2}F_{1}}$.
 
\subsection{Definition of concentration}
 
The mass enclosed within a region $r_{\rm \Delta}$ may be written in
terms of the critical density of the Universe,
 
\begin{equation}
\rho_{\rm crit}(z_{\rm f}) = 3H(z_{\rm f})^{2}/8 \pi G,
\label{eqn:rhocrit}
\end{equation}
where $z_{\rm f}$ is the cluster formation redshift, assumed equal to
the observed redshift, such that
 
\begin{equation}
M_{\Delta}=\frac{4}{3} \pi r_{\Delta}^{3} \rho_{\rm crit}  \Delta,
\label{eqn:rdelta}
\end{equation}
where $\Delta$ is the density contrast. The Hubble parameter, $H(z) =
100 h E(z) \kmpspMpc$, is derived from the Friedmann equation.  In a
flat Universe \(E(z)^{2} = \Omega_{\rm m}(1+z)^{3}+\Omega_{\Lambda}\).
 
The concentration, $c_{\Delta}$, is defined such that
\(c_{\Delta}=r_{\Delta}/r_{\rm s}\). Using this definition the
normalization of the density profiles, $\rho_{0}$, may be written as
the product of the critical density and a dimensionless parameter,
$\delta_{c}$.  Substituting $x_{\Delta} = r_{\Delta}/r_{\rm s} =
c_{\Delta}$ into the mass formulae above (Equations~\ref{eqn:darkmass}
and ~\ref{eqn:darkgen}) at $r = r_{\Delta}$ and setting them equal to
Equation~\ref{eqn:rdelta} yields the following expressions for
$\delta_{c}$
 
\begin{equation}
\delta_{c}  =
\begin{cases}
\frac{\Delta}{3}\frac{c_{\Delta}^{3}}{\mathrm{ln}(1+c_{\Delta})-c_{\Delta}/(1+c_{\Delta})}& \text{for NFW}\\
\frac{\Delta}{2}\frac{c_{\Delta}^{3}}{\mathrm{ln}(1+c_{\Delta}^{3/2})}
& \text{for M99}\\
\frac{\Delta}{3}\frac{c_{\Delta}^{\alpha} \alpha_{3}}{{_{2}F_{1}}(\alpha_{3}/\gamma,(\beta-\alpha)/\gamma;(\alpha_{3}+\gamma)/\gamma;-c_{\Delta}^{\gamma})}
 & \text{for M${_{2}F_{1}}$}
\end{cases}
\label{eqn:deltac}
\end{equation}

\subsection{Definition of the virial radius}
The virial radius separates the region where the cluster is in
hydrostatic equilibrium from where matter is still infalling.  For
\(\Omega_{\rm m}+\Omega_{\rm \Lambda} = 1\), $r_{\Delta} = r_{\rm
  vir}$ if \(\Delta = \Delta_{\rm vir} = 178\Omega_{\rm m}^{0.45}\)
\citep{lacey&cole1993,ekeetal1996,ekeetal1998}. For $\Omega_{\rm m}$ =
0.3, $ \Delta_{\rm vir} = 104$. In an Einstein-de Sitter Universe,
$\Omega_{\rm m}$ = 1.0 and $ \Delta_{\rm vir} = 178$. A density
contrast of $\Delta$ = 200 (or $\Delta$ = 178) is still often used in
the literature and $r_{\rm vir}$ and $r_{200}$ (or $r_{178}$) written
interchangeably. For purposes of comparison with other work in the
literature we adopt $\Delta = 200$.

\section{Modelling the observed mass profiles}
CDM simulations make two robust predictions concerning the
distribution of mass in dark matter halos: the shape of the mass
profile differs strongly from a power-law, and the power-law slope of
the inner density profile lies in the range
$1.0\lesssim\alpha\lesssim1.5$ \citep[see, for
example,][]{navarroetal2004}. These properties are investigated for
the observed mass profiles below.
 
\subsection{General shape of the mass profile}
\begin{figure*}
\includegraphics[angle=0,width=0.95\columnwidth]{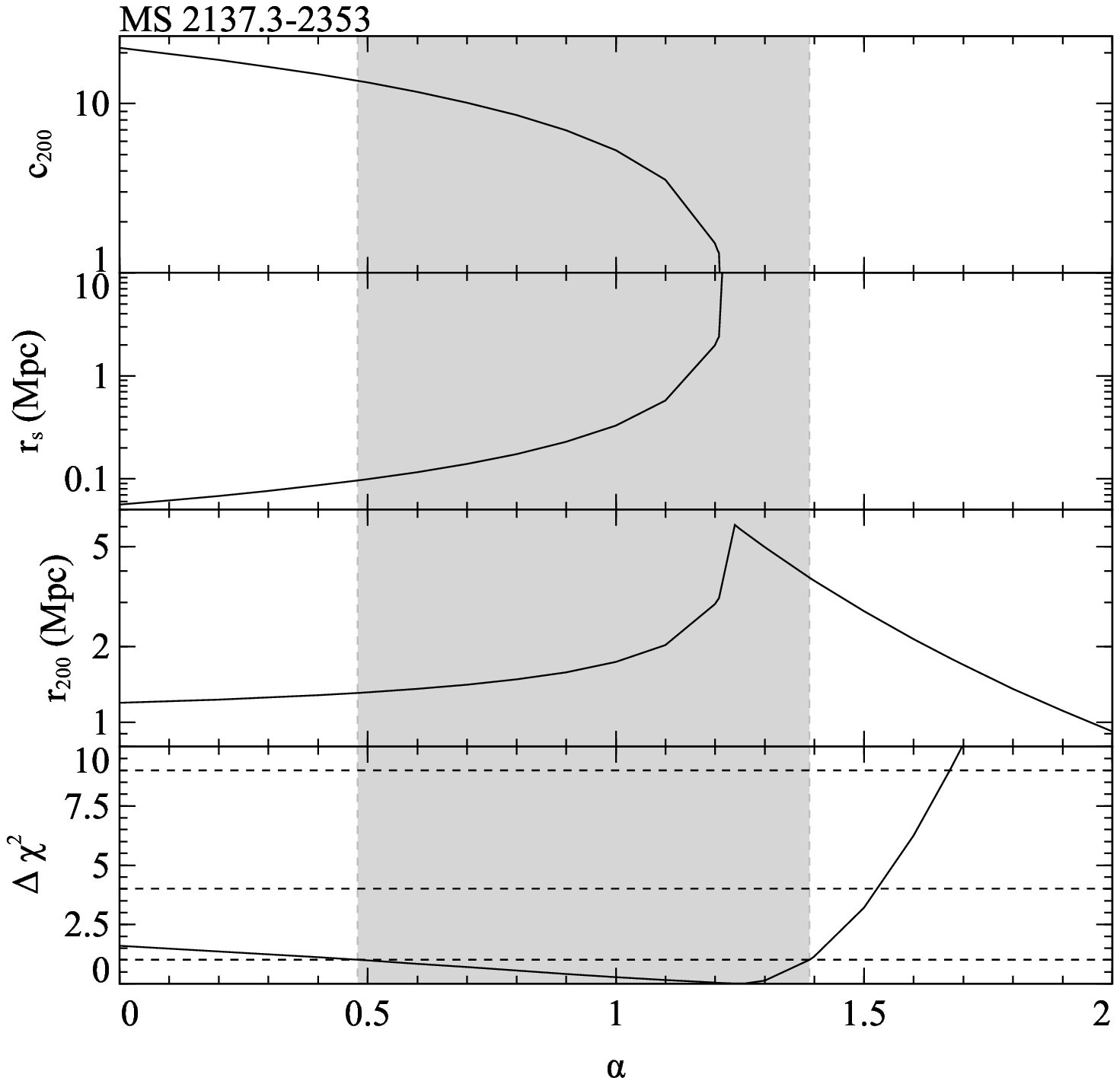}
\vspace{1.0mm}
\includegraphics[angle=0,width=0.95\columnwidth]{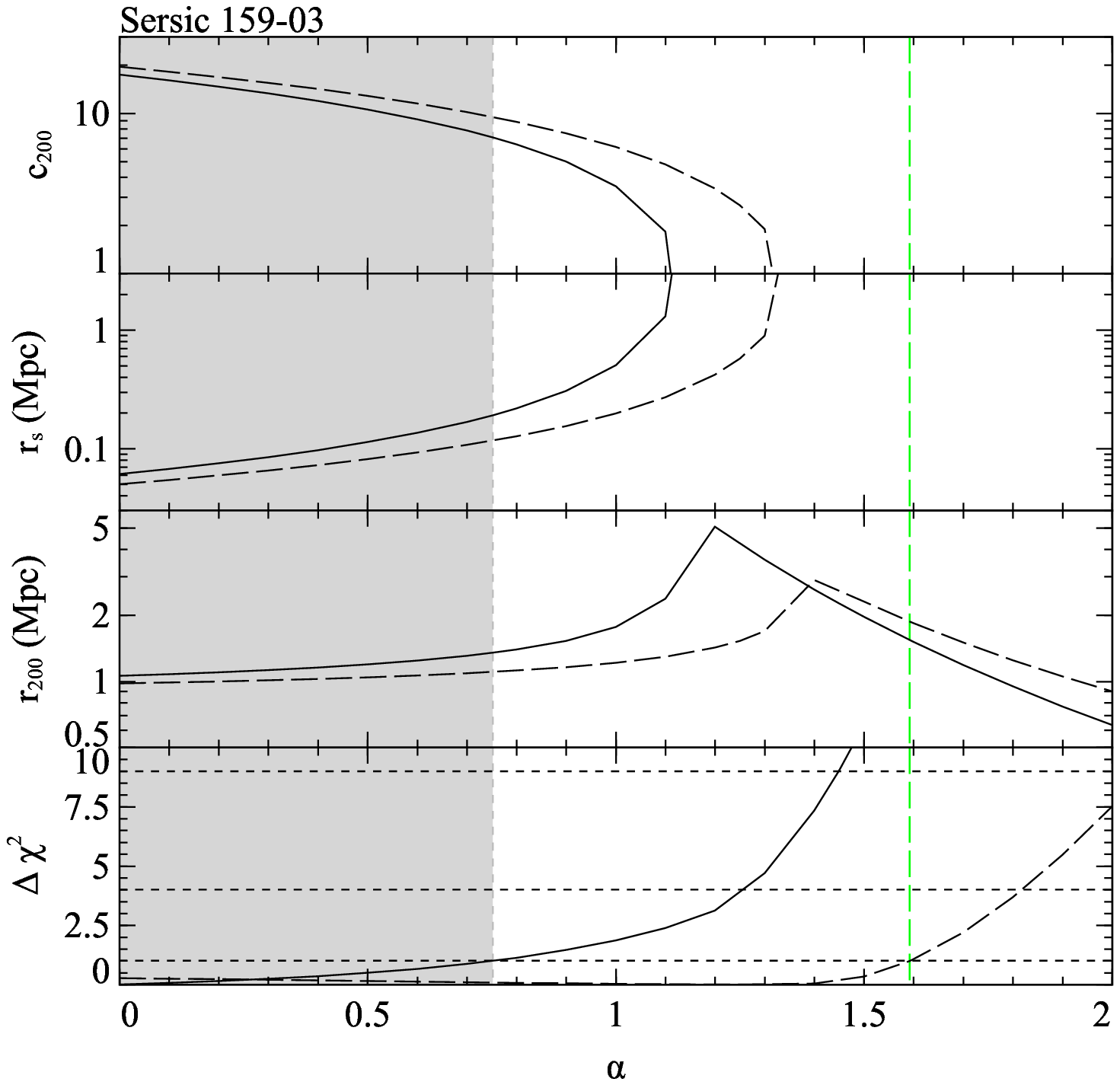}
\caption{Plots showing variation in concentration, scale radius,
  virial radius and $\Delta \chi^{2}$ ($= \chi^{2} -
  \chi^{2}_{\rm{min}}$, where $\chi^{2}_{\rm{min}}$ is the minimum
  chi-square) for values of the inner slope $\alpha$ between 0 and 2.
  The dotted horizontal lines in the bottom panel show the 1$\sigma$
  (68.3 per cent), 2$\sigma$ (95.4 per cent) and 3$\sigma$ (99.7 per
  cent) confidence levels for alpha.  The grey shaded regions show the
  parameter values which lie within the one-sigma confidence interval.
  Note that when the concentration drops below 1.0 the scale radius is
  larger than the virial radius and the profile tends to a power-law.
  Fit parameters determined with the inner 30 kpc excluded are shown
  with dashed lines, and the one-sigma confidence limits shown by
  green dashed lines. (If only one green dashed line is shown then the
  lower limit on alpha is less than zero).}
\label{fig:chi}
\end{figure*}

The mass profiles in Fig.~\ref{fig:mass} are fit with the generalized
mass model in Equation~\ref{eqn:darkgen}, with $\beta$ and $\gamma$
fixed at 1.0.  Fig.~\ref{fig:chi} shows the variation in concentration
parameter, scale radius, virial radius ($r_{200}$) and chi-square with
inner slope as $\alpha$ is stepped through values between 0.0 and 2.0
and the fit is minimized with respect to $c$ and $r_{\rm s}$. (Note
that the scale radius and concentration are highly negatively
correlated). The virial radius is computed by substituting
\(c_{\Delta}=r_{\Delta}/r_{\rm s}\) into Equation~\ref{eqn:deltac}.

\begin{figure*}
\centering
\includegraphics[angle=0,width=0.95\columnwidth]{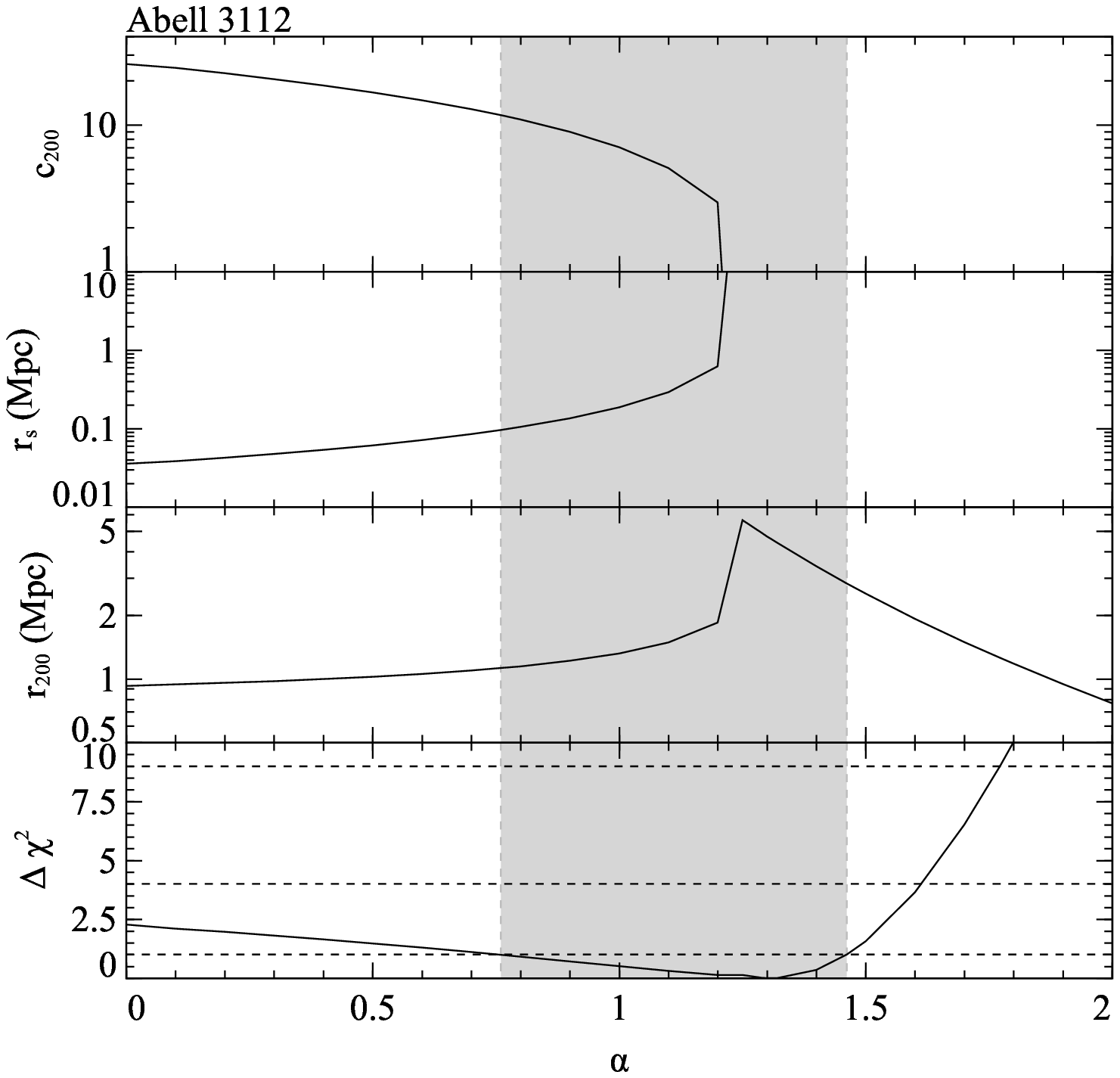}
\vspace{1.0mm}
\includegraphics[angle=0,width=0.95\columnwidth]{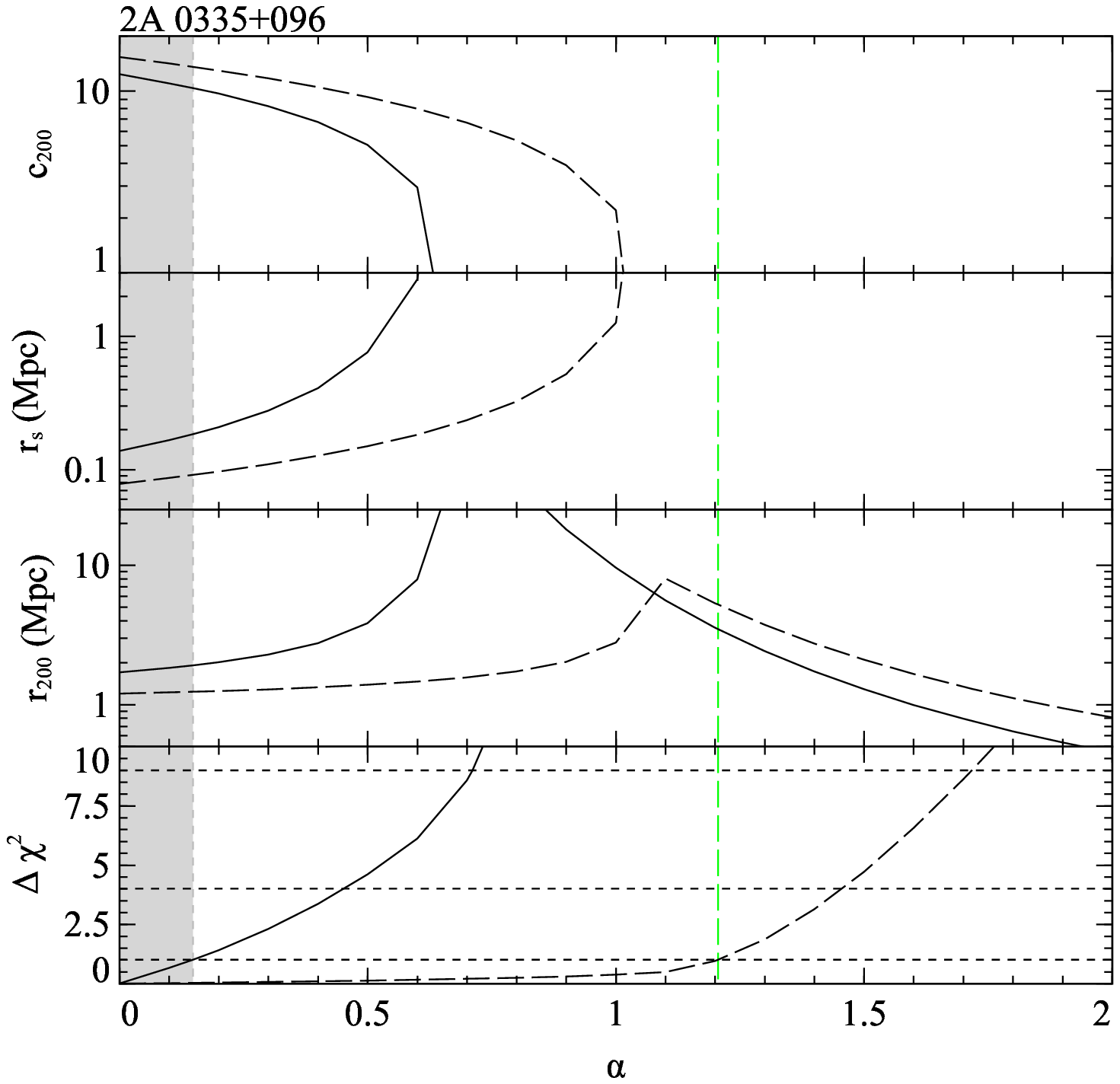}
\vspace{1.0mm}
\includegraphics[angle=0,width=0.95\columnwidth]{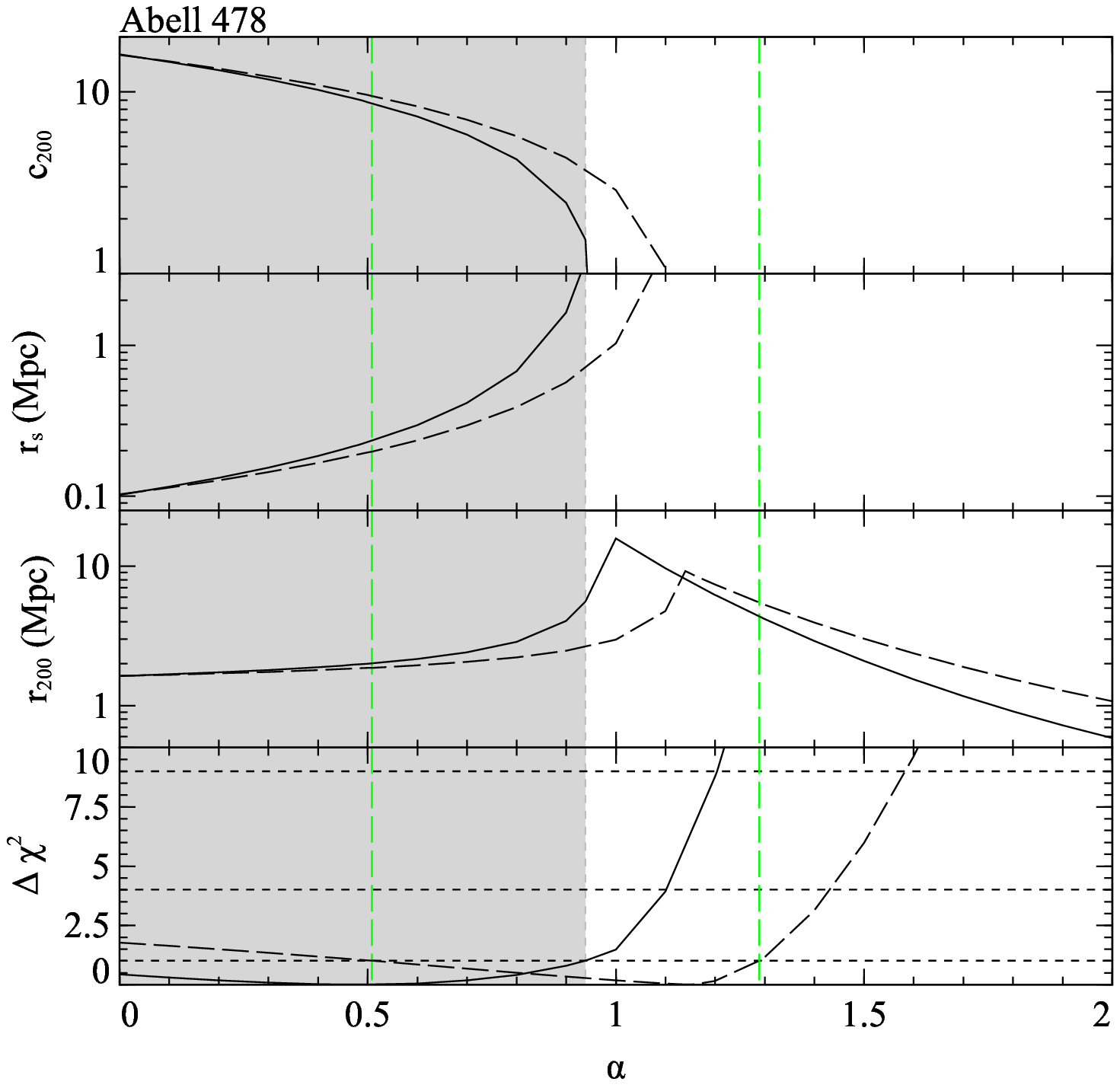}
\vspace{1.0mm}
\includegraphics[angle=0,width=0.95\columnwidth]{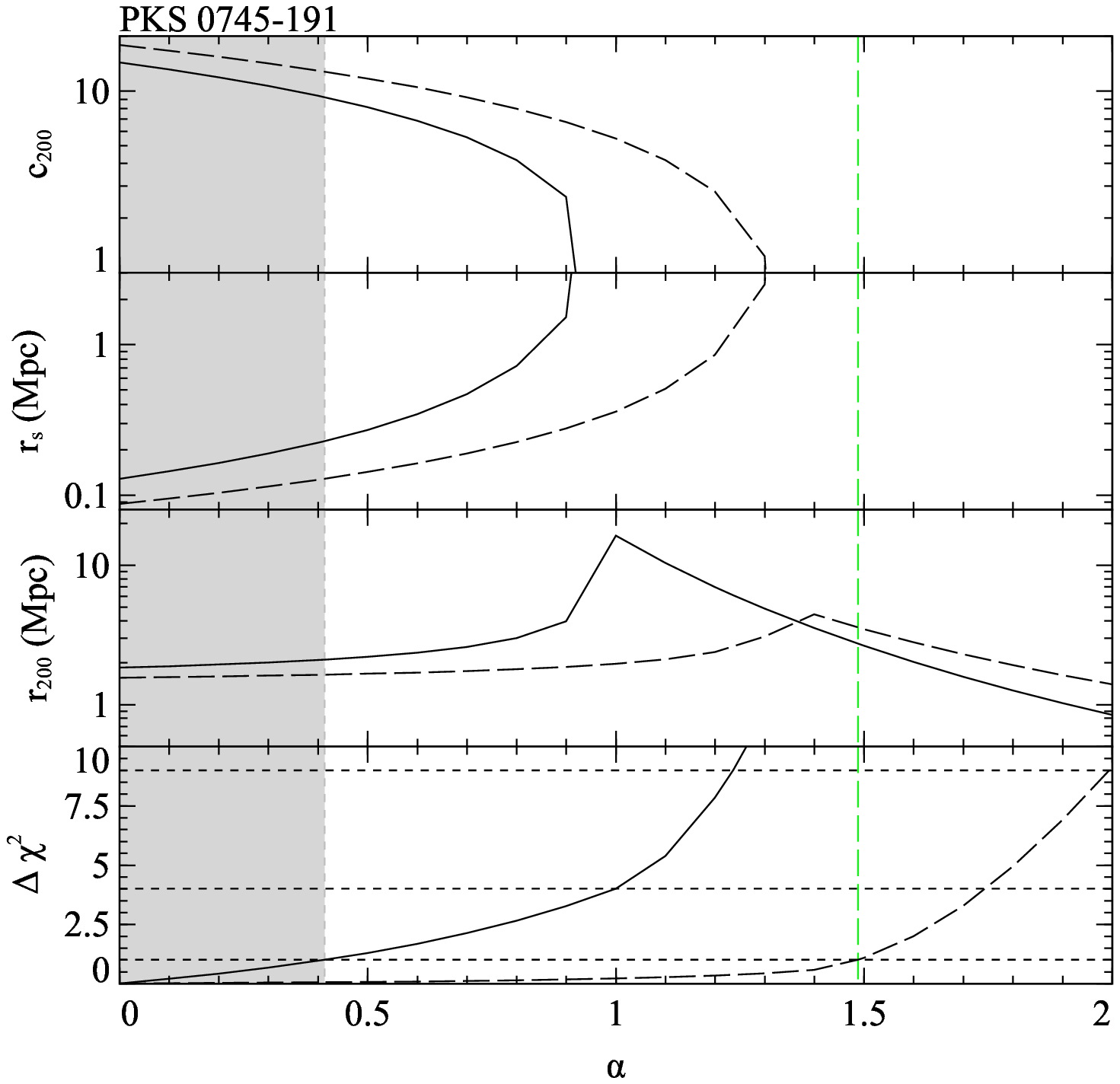}
\vspace{1.0mm}
\includegraphics[angle=0,width=0.95\columnwidth]{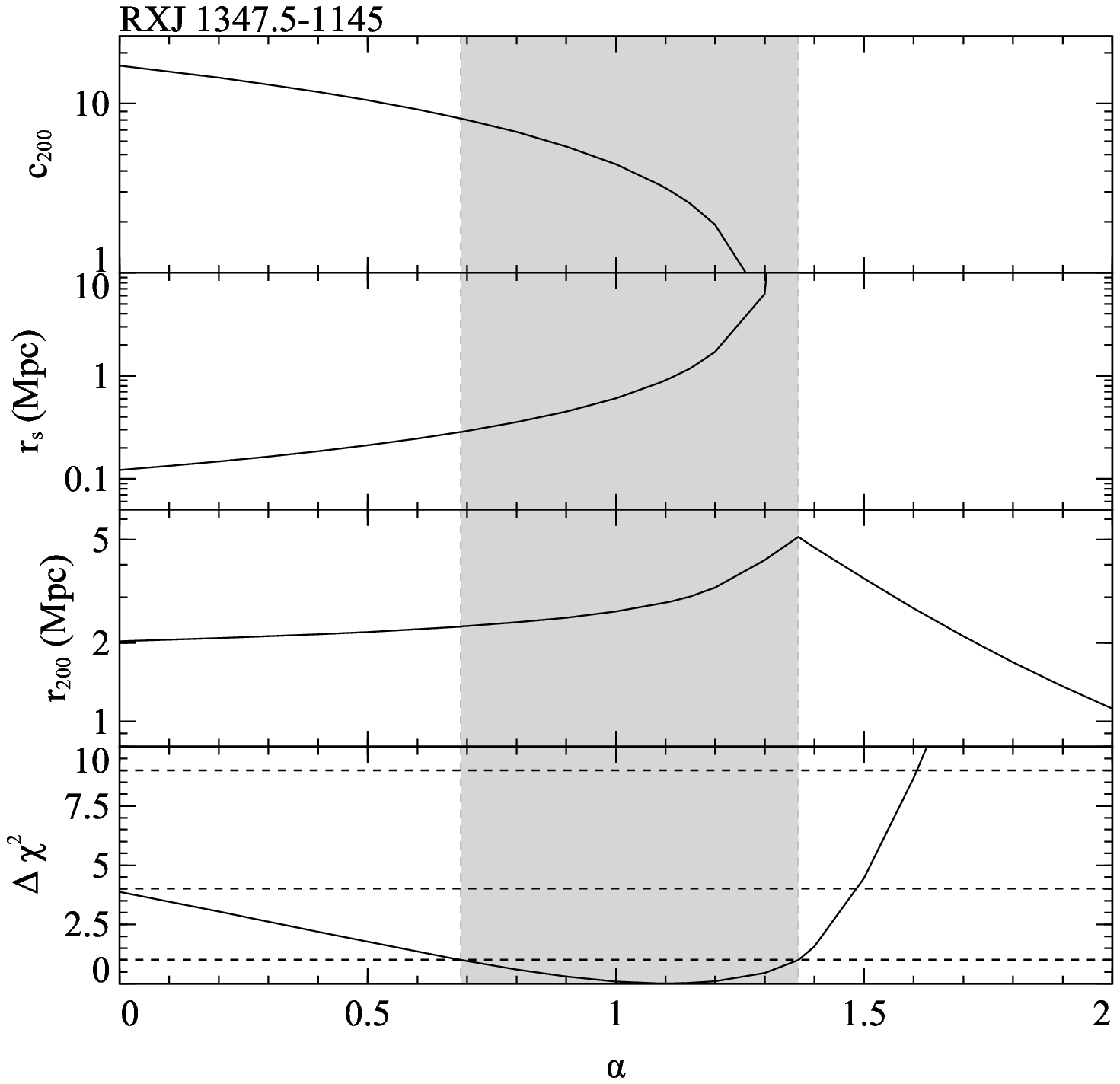}
\vspace{1.0mm}
\includegraphics[angle=0,width=0.95\columnwidth]{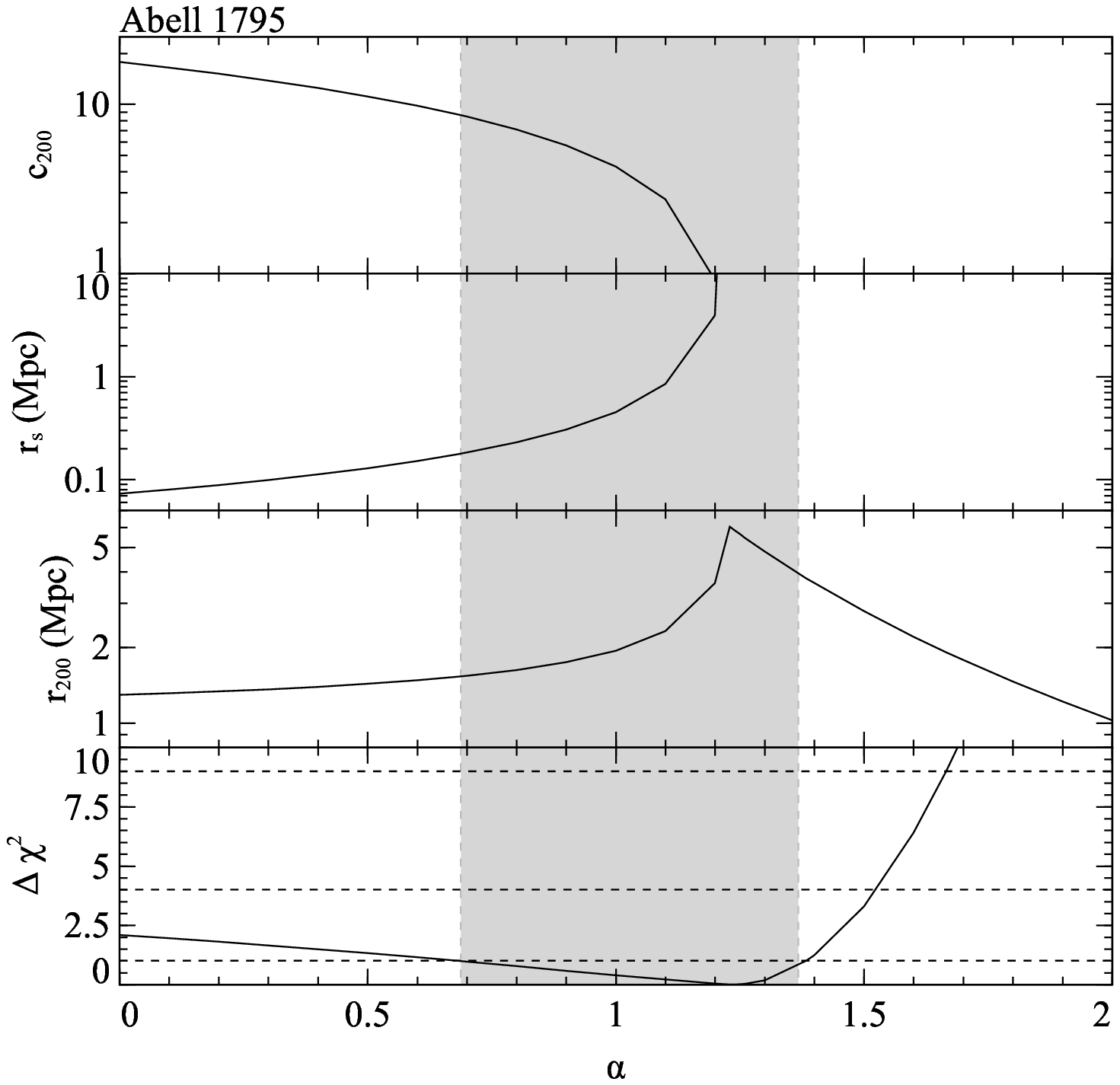}
\vspace{1.0mm}
\end{figure*}
\begin{figure*}
\centering
\includegraphics[angle=0,width=0.95\columnwidth]{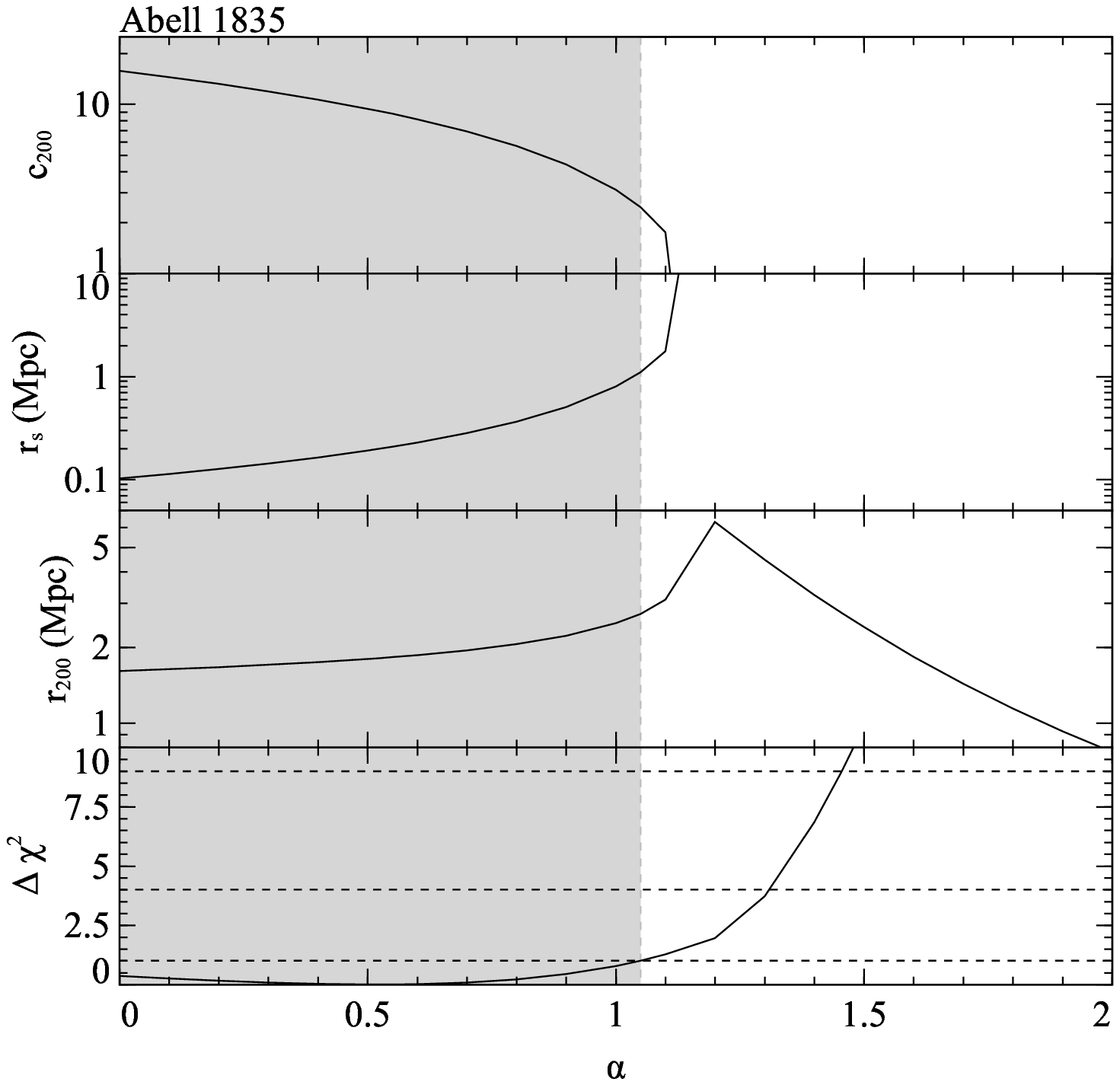}
\vspace{1.0mm}
\includegraphics[angle=0,width=0.95\columnwidth]{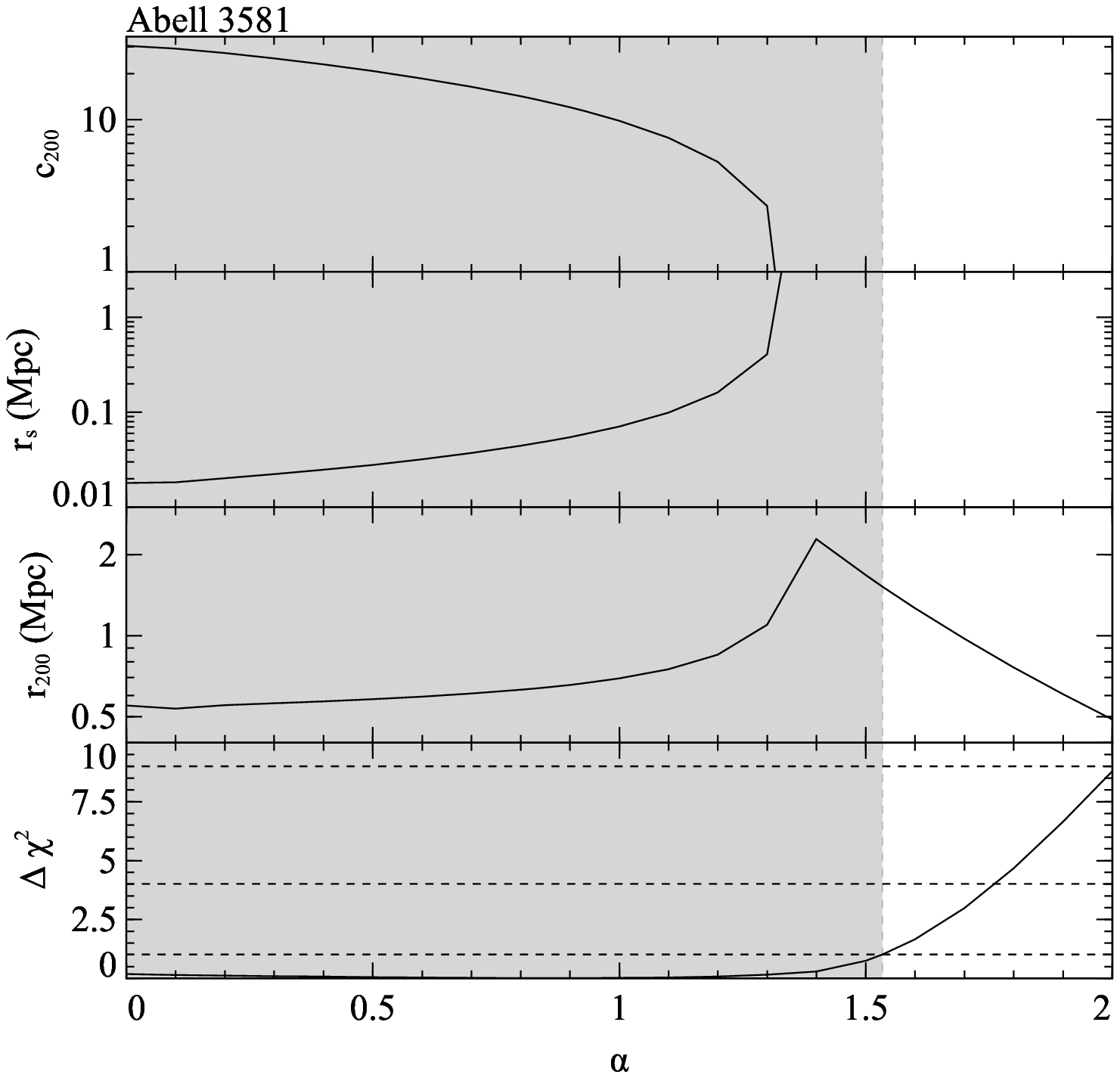}
\vspace{1.0mm}
\includegraphics[angle=0,width=0.95\columnwidth]{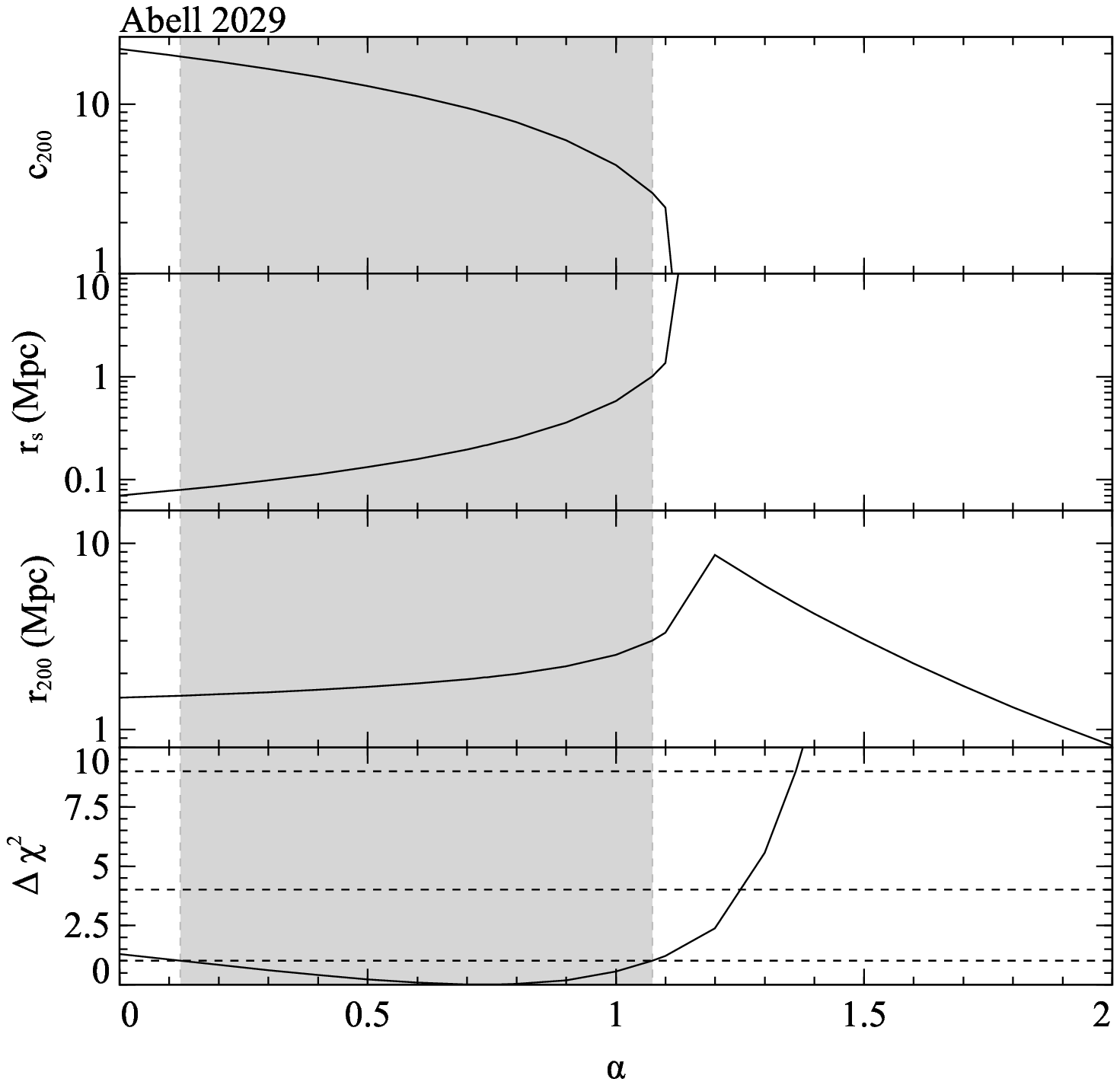}
\vspace{1.0mm}
\includegraphics[angle=0,width=0.95\columnwidth]{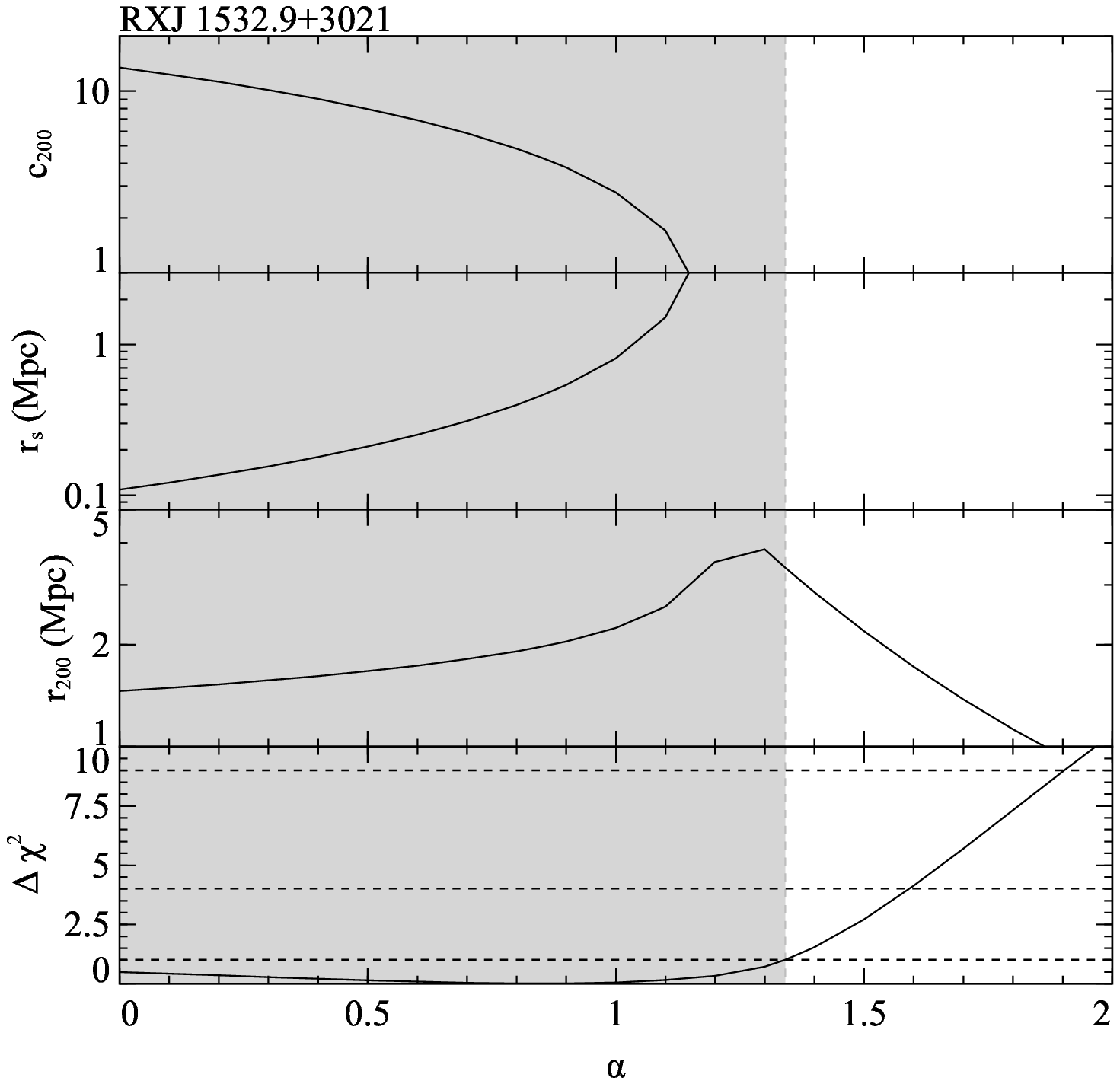}
\vspace{1.0mm}
\end{figure*}

If the best-fit concentration is less than 1.0 for a particular value
of $\alpha$, then the mass model does not agree with CDM predictions
i.e. the characteristic radius, $r_{\rm s}$, at which the profile
`turns-over' is greater than the virial radius. In
Table~\ref{tab:innerslope} we show the best-fit parameters for the
$M_{2}F_{1}$ and power-law ($M_{\rm pow} \propto r^{3-\alpha}$)
models. When the best-fit concentration for the $M_{2}F_{1}$ model is
$c \ll 1.0$ the model tends to a power-law. We refer to the best-fit
model as a power-law when the best-fit concentration is less then 1.0.
Although not strictly true, the description serves as a useful
distinction between profiles which are consistent ($c>1$) and those
which are inconsistent ($c<1$) with CDM model predictions.

\begin{table*}
\centering
\begin{tabular}{lccccccc}
\hline
Cluster & \multicolumn{5}{c}{M${_{2}F_{1}}$ model} &
  \multicolumn{2}{c}{Power-law model} \\
& $\alpha$ & $c_{200}$ & $r_{\rm s}$ & $r_{200}$ & $\chi^{2}$(dof)&
$\alpha$ & $\chi^{2}$(dof) \\
& & & (Mpc) & (Mpc) & & & \\
\hline
Abell~3112 & 1.31$^{+0.15}_{-0.55}$ &-- &-- & 4.6 & 0.8(3)&
1.31$^{+0.15}_{-0.16}$& 0.8(4) \\
\\
2A~0335$+$096 & 0.00$^{+0.15}_{-0.00}$ & 12 & 0.14 & 1.7 & 6.5(2)& 0.55$^{+0.09}_{-0.09}$& 12.3(3)\\
& 0.00$^{+1.21}_{-0.00}$ &15  & 0.08 & 1.2 & 2.3(1)&
1.05$^{+0.21}_{-0.19}$ & 0.7(2) \\
\\
Abell~478 & 0.49$^{+0.45}_{-0.49}$ &  9.0 & 0.22 & 2.0 & 6.7(5) & 0.94$^{+0.10}_{-0.10}$& 7.8(6)\\
& 1.14$^{+0.15}_{-0.63}$  & --  & -- & 9.2 & 5.4(4)&
1.14$^{+0.15}_{-0.16}$& 5.4(5) \\
\\
PKS~0745$-$191 & 0.00$^{+0.41}_{-0.00}$ & 14 & 0.13  & 1.8  &  2.5(2)&
0.91$^{+0.15}_{-0.16}$& 6.1(3) \\
& 0.00$^{+1.49}_{-0.00}$ & 18 & 0.09 & 1.6 & 0.4(1)&
1.32$^{+0.22}_{-0.22}$& 0.9(2) \\
\\
RXJ~1347.5$-$1145 & 1.10$^{+0.27}_{-0.41}$ & 3.2  & 0.90 & 2.9 & 3.0(1)& 1.30$^{+0.10}_{-0.10}$
& 3.4(2) \\
\\
Abell~1795 & 1.24$^{+0.15}_{-0.56}$ & -- & -- & 5.8 & 2.6(2)&
1.24$^{+0.15}_{-0.16}$&  2.6(3) \\
\\
Abell~1835 & 0.52$^{+0.53}_{-0.52}$ & 9.2  & 0.20  & 1.8  & 5.9(3)&
1.14$^{+0.11}_{-0.11}$& 7.5(4)\\
\\
Abell~3581 & 0.88$^{+0.65}_{-0.88}$ & 12 & 0.05 & 0.7 & 1.3(3)&
1.34$^{+0.22}_{-0.24}$& 1.5(4)\\
\\
Abell~2029 & 0.74$^{+0.34}_{-0.61}$  & 8.9 & 0.21 & 1.9 & 2.4(4) &
1.13$^{+0.09}_{-0.09}$& 4.1(5)\\
\\
RXJ~1532.9$+$3021 & 0.85$^{+0.49}_{-0.85}$ & 4.3  & 0.46 & 2.0 & 1.2(2)
& 1.18$^{+0.20}_{-0.19}$& 1.6(3)\\
\\
MS~2137.3$-$2352 & 1.25$^{+0.14}_{-0.77}$ & -- & -- & 5.9 & 0.2(1) &
1.25$^{+0.14}_{-0.15}$ & 0.2(2)\\
\\
Sersic~159$-$03 & 0.00$^{+0.75}_{-0.00}$ & 17 & 0.06 & 1.1 & 3.4(3)&
1.09$^{+0.14}_{-0.15}$& 6.0(4)\\
& 1.20$^{+0.39}_{-1.20}$ & 3.4 & 0.42 & 1.4 & 2.5(2)&
1.38$^{+0.21}_{-0.21}$& 2.5(3) \\
\hline
\end{tabular}
\caption{Best-fit slope $\alpha$ and corresponding chi-square for the
  M${_{2}F_{1}}$ and
  power-law models. The second row for 2A~0335$+$096, Abell~478, PKS~0745$-$191 and Sersic~159$-$03 shows the best-fit with the inner 30 kpc excluded.
  The best-fit concentration, scale radius and virial radius are shown for
  the M${_{2}F_{1}}$ model. The concentration and scale radius
  are not shown for clusters where the best-fit tends to a power-law
  (i.e. $c<<1.0$).}
\label{tab:innerslope}
\end{table*}

With $\alpha$ a free parameter in the fits, the concentration is
consistent with both $c > 1$ and $c < 1$ within the one sigma
uncertainties for half the objects analysed (Abell~3112,
RXJ~1347.5$-$1145, Abell~1795, Abell~3581, RXJ~1532.9$+$3021 and
MS~2137.3$-$2353; see shaded regions in Fig.~\ref{fig:chi}). In four
clusters (2A~0335$+$096, Abell~478, PKS~0745$-$191 and
Sersic~159$-$03) the best-fit concentration is greater than 1.0 at the
68 per cent confidence level; although the best-fit inner slope in
these objects is shallower than predicted by CDM simulations.
Abell~1835 and Abell~2029 are the only two objects for which a
power-law is ruled out at the 1$\sigma$ level \emph{and} the inner
slope is consistent with the CDM model.                                                                                       
\subsection{Inner slope measurements}
 
\begin{figure*}
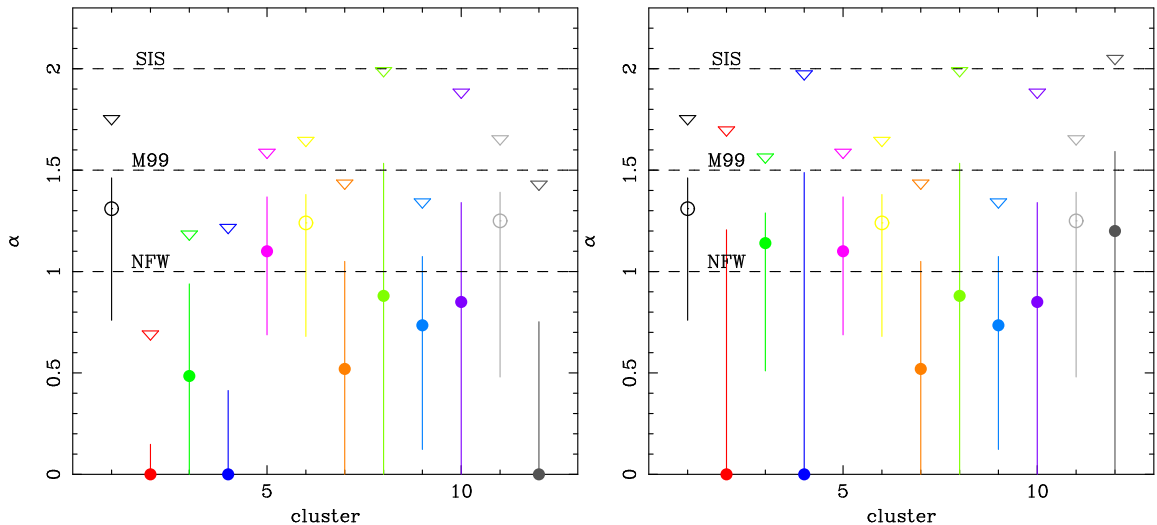

\centering
  \includegraphics[angle=-90,width=0.9\columnwidth]{alpha_all_apr05.ps}
  \includegraphics[angle=-90,width=0.9\columnwidth]{alpha_all_outer_apr05.ps}
  \caption{Left: best-fit inner slope. Right: best-fit inner slope
    with inner 30 kpc removed in 2A~0335$+$096, Abell~478,
    PKS~0745$-$191 and Sersic~159$-$03. The error bars show the 1
    sigma uncertainties and the downward arrows the 3 sigma upper
    limits. An open circle is used to indicate clusters with a
    best-fit concentration less than 1.0. The clusters are plotted
    from left to right in the same order as they are listed (from top
    to bottom) in Table~\ref{clusters}.}
\label{fig:alpha}
\end{figure*}

The best-fit inner slope and 1$\sigma$ uncertainties are plotted for
each object in Fig.~\ref{fig:alpha}. Also shown are the 3$\sigma$
upper limits on the slope. Clusters with a best-fit concentration less
than 1.0 are indicated. It is clear from the plot that 8 out of the 12
clusters analysed are consistent with the CDM model.  Furthermore, the
points appear to scatter around $\alpha=1.0$, rather than
$\alpha=1.5$. An inner logarithmic slope of $\alpha=2$ is ruled out at
the 3$\sigma$ level in all objects except Abell~3581; the mass
profiles would therefore seem to rule out the single isothermal sphere
(SIS) model which has been widely used in the literature
\citep[e.g.][]{binney&tremaine1987}. We note that there is no obvious
correlation between the cluster mass and the best-fit inner slope,
although the uncertainties are too large to justify any correlation
tests on the data.
                                                                                         
The four clusters which are inconsistent with an inner slope of 1.0 at
the 1$\sigma$ level are 2A~0335$+$096, Abell~478, PKS~0745$-$191 and
Sersic~159$-$03, with Abell~478, Sersic~159$-$03 and PKS~0745$-$191
consistent with $\alpha=1.0$ at the 2 sigma level. Only 2A~0335$+$096
disagrees with CDM predictions with 99.7 per cent confidence. We note
that this is also the object for which the spectral model used
provides the poorest fit (see Section 3.5) and, as discussed below,
the temperature distribution in the centre is highly asymmetric
between the northern and southern sectors of the cluster. The presence
of an asymmetric temperature distribution in 2A~0335$+$096 has been
shown previously by \citet{mazzottaetal2003}.

In a recent study, \citet{katayama&hayashida2004} find that six out of
the twenty clusters in their sample (including Abell~478,
PKS~0745$-$191 and 2A~0335$+$096) have a slope lower than unity at the
90 per cent confidence level. The authors interpret this result as
being in agreement with the `flat core problem' found in observations
of low mass objects.
 
The method used for computing the mass profiles relies on the
assumptions of hydrostatic equilibrium and spherical symmetry. Whether
or not the former condition is met is difficult to ascertain.
\citet{katayama&hayashida2004} touch on this problem by highlighting
objects which exhibit obvious structure in their cores, suggesting
that this may indicate a break from hydrostatic equilibrium, as well
as the possibility of affecting the ambient temperature and density
profile measurements. They conclude that some objects have flat inner
slopes (Abell~478, PKS~0745$-$191, ZW~3146), even when those with
central structure (2A~0335$+$096 and Abell~2597) are removed.
 
In Fig~\ref{fig:alpha} we plot the best-fit inner density slopes when
the inner $\sim30 \kpc$ is removed from the fits to 2A~0335$+$096,
Abell~478, PKS~0745$-$191 and Sersic~159$-$03. It is clear that it is
the innermost data point in each of these clusters which is forcing
the best-fit to be flat. The question is `how far can we trust data
from the central 30 kpc?'.  The challenge will be to determine the
extent of any non-thermal pressure support in the central regions of
these objects.
 
\subsection{Symmetry of the temperature distribution}
One clue that the gas is disturbed and not in hydrostatic equilibrium
may be suggested by asymmetries in the temperature distribution. Here
we investigate whether there is any correlation between the degree of
symmetry in the temperature distribution and the inner slope of the
mass profile.

\begin{figure*}
\centering
\includegraphics[angle=-90,width=0.95\columnwidth]{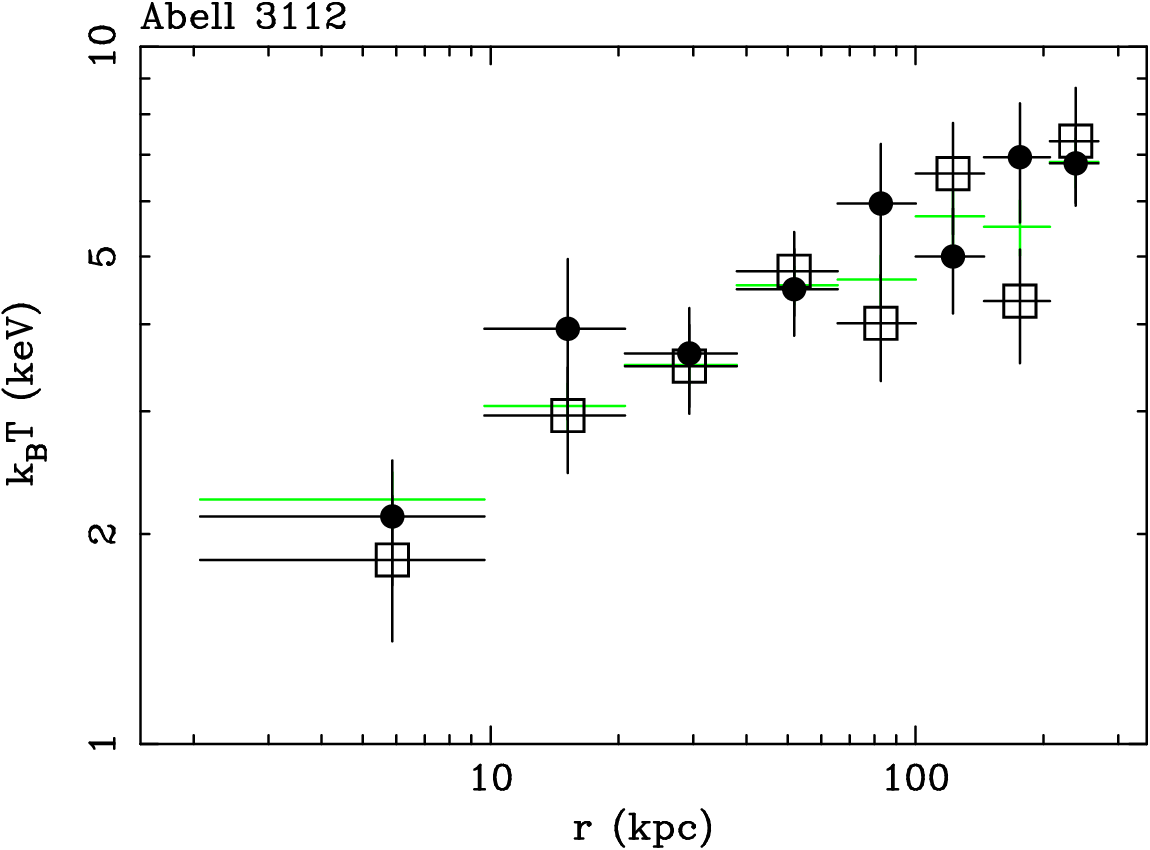}
\vspace{1.0mm}
\includegraphics[angle=-90,width=0.95\columnwidth]{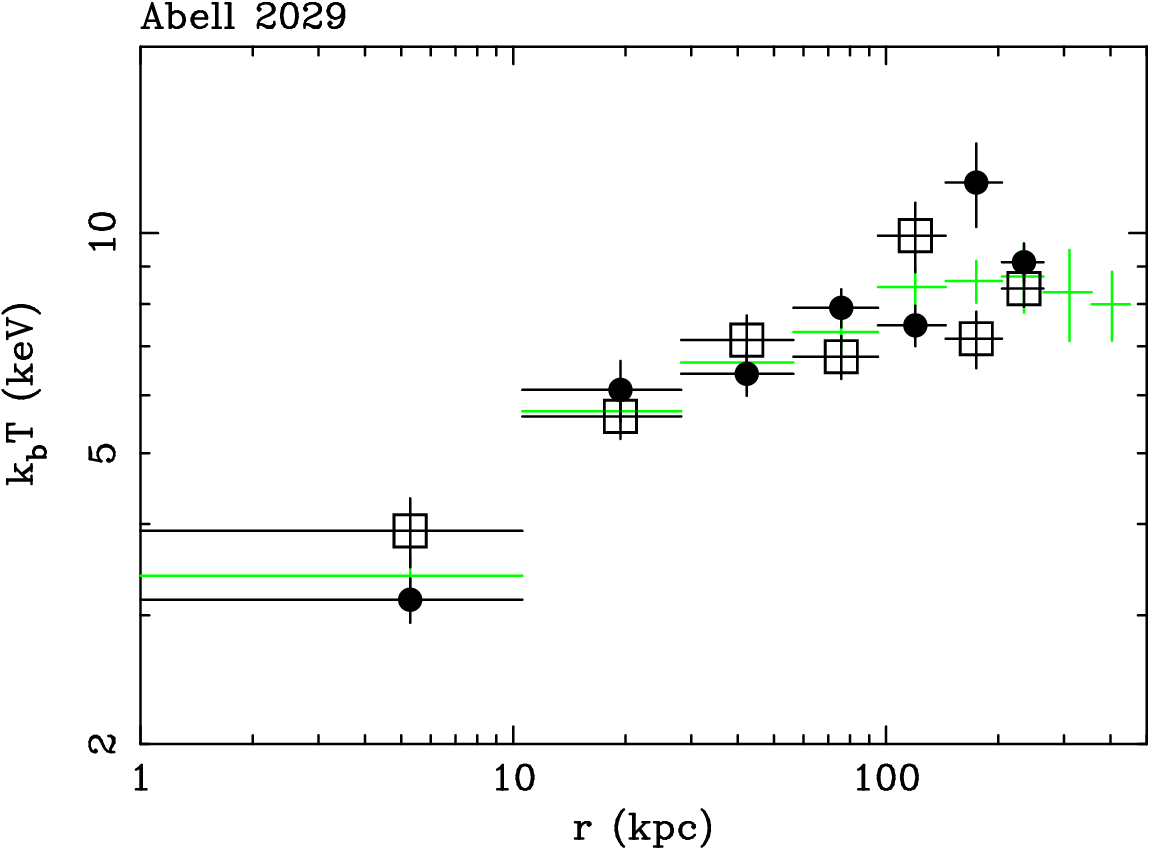}
\vspace{1.0mm}
\includegraphics[angle=-90,width=0.95\columnwidth]{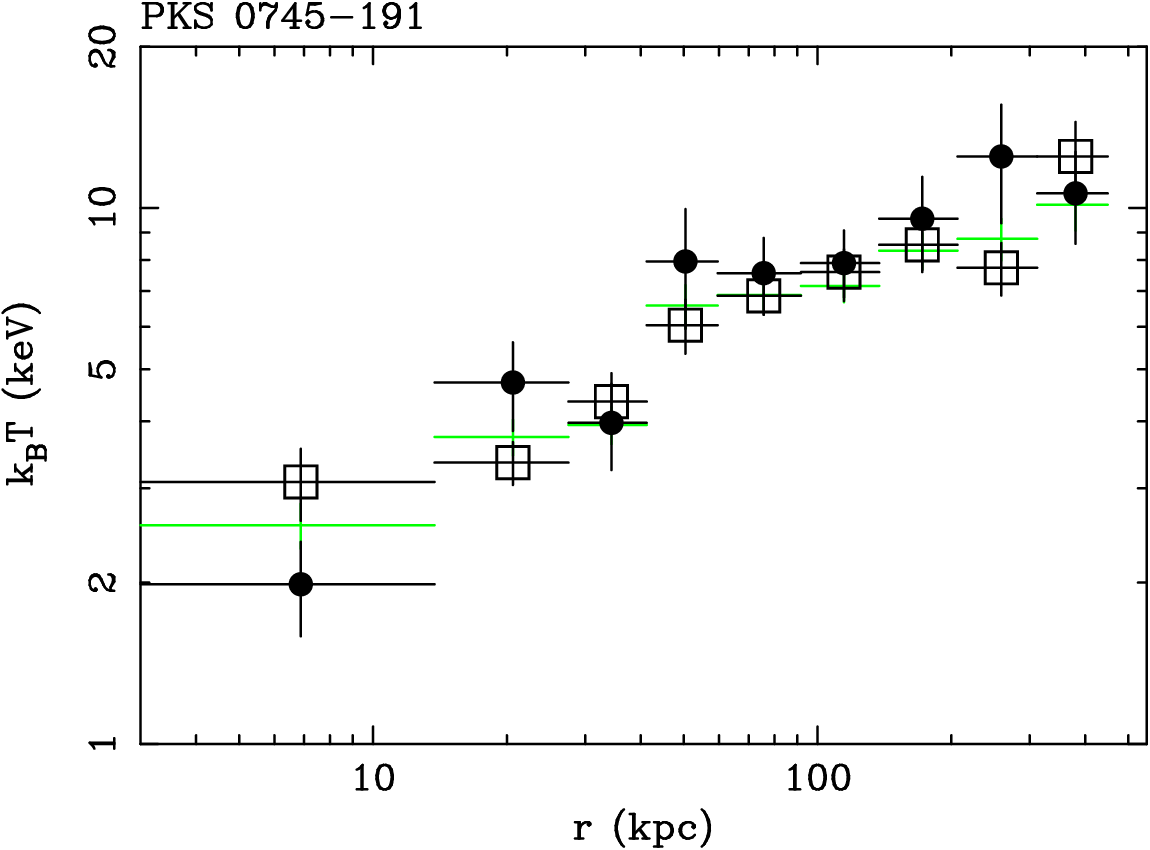}
\vspace{1.0mm}
\includegraphics[angle=-90,width=0.95\columnwidth]{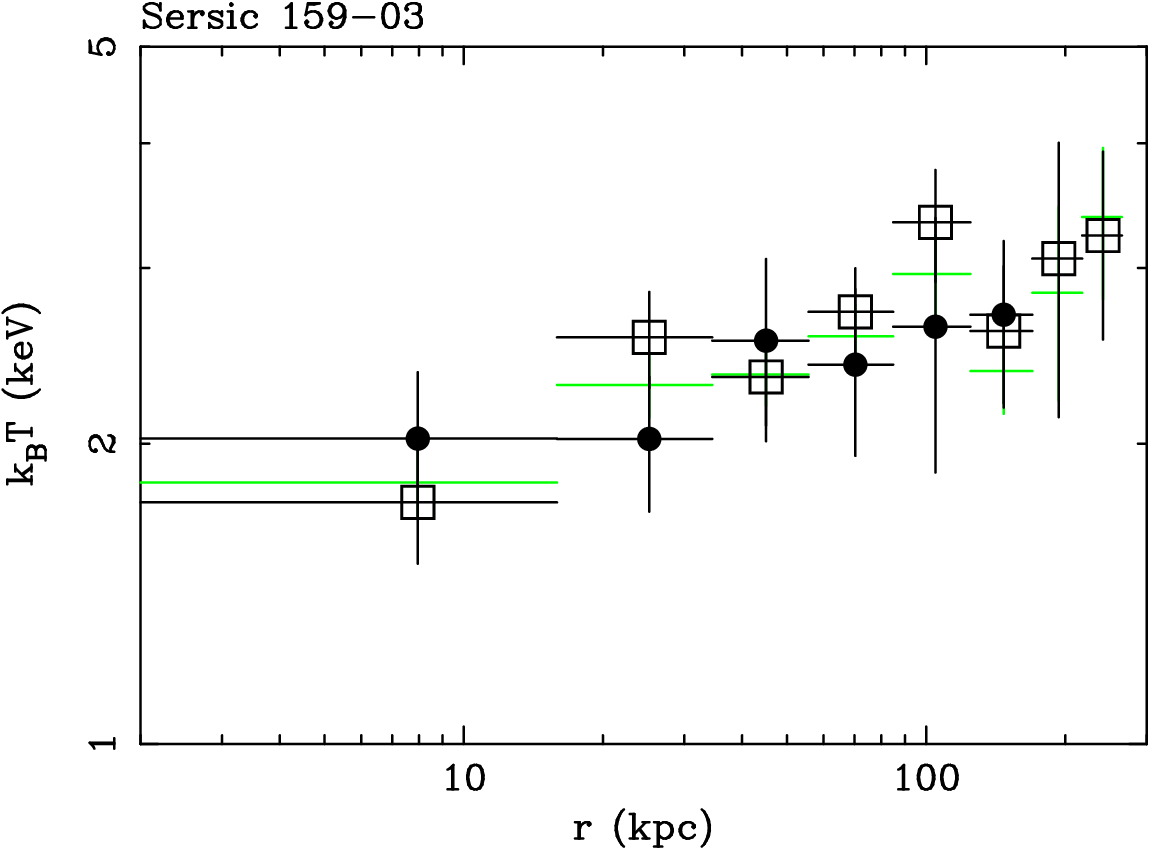}
\vspace{1.0mm}
\includegraphics[angle=-90,width=0.95\columnwidth]{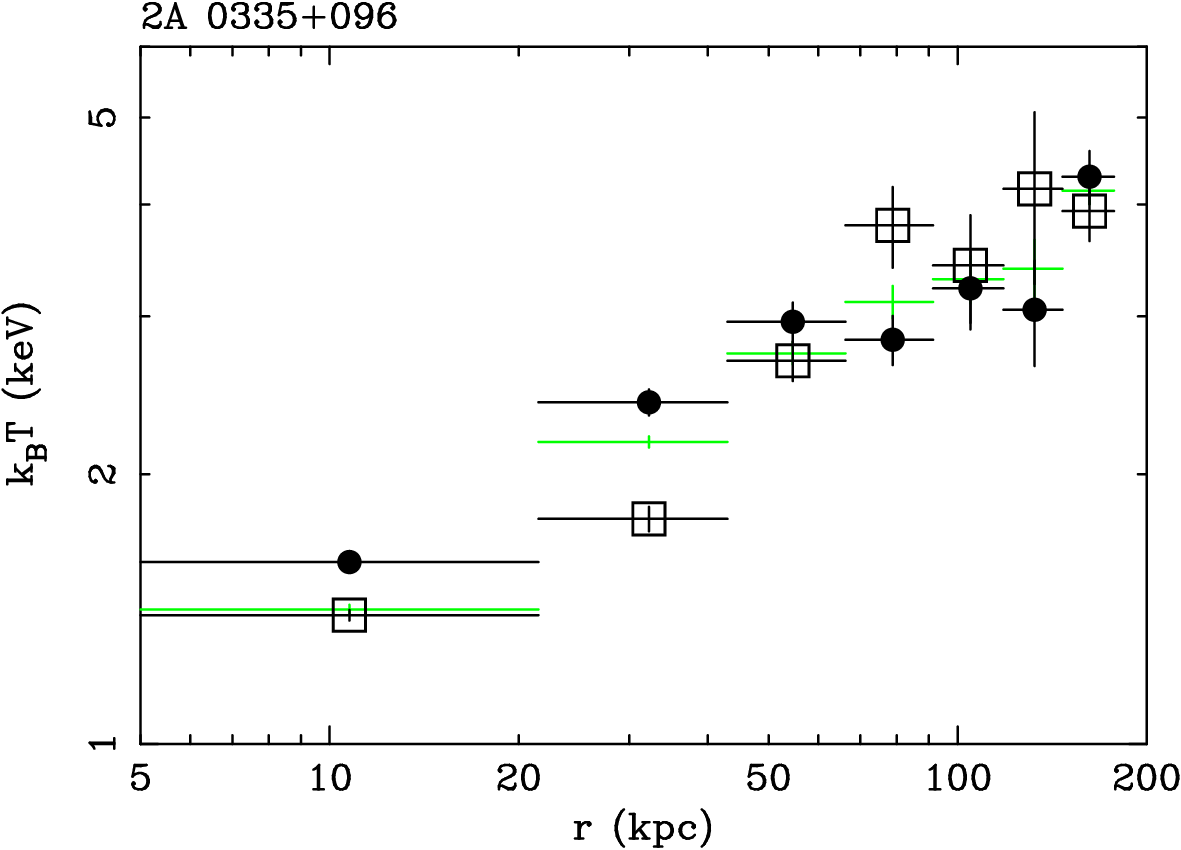}
\vspace{1.0mm}
\includegraphics[angle=-90,width=0.95\columnwidth]{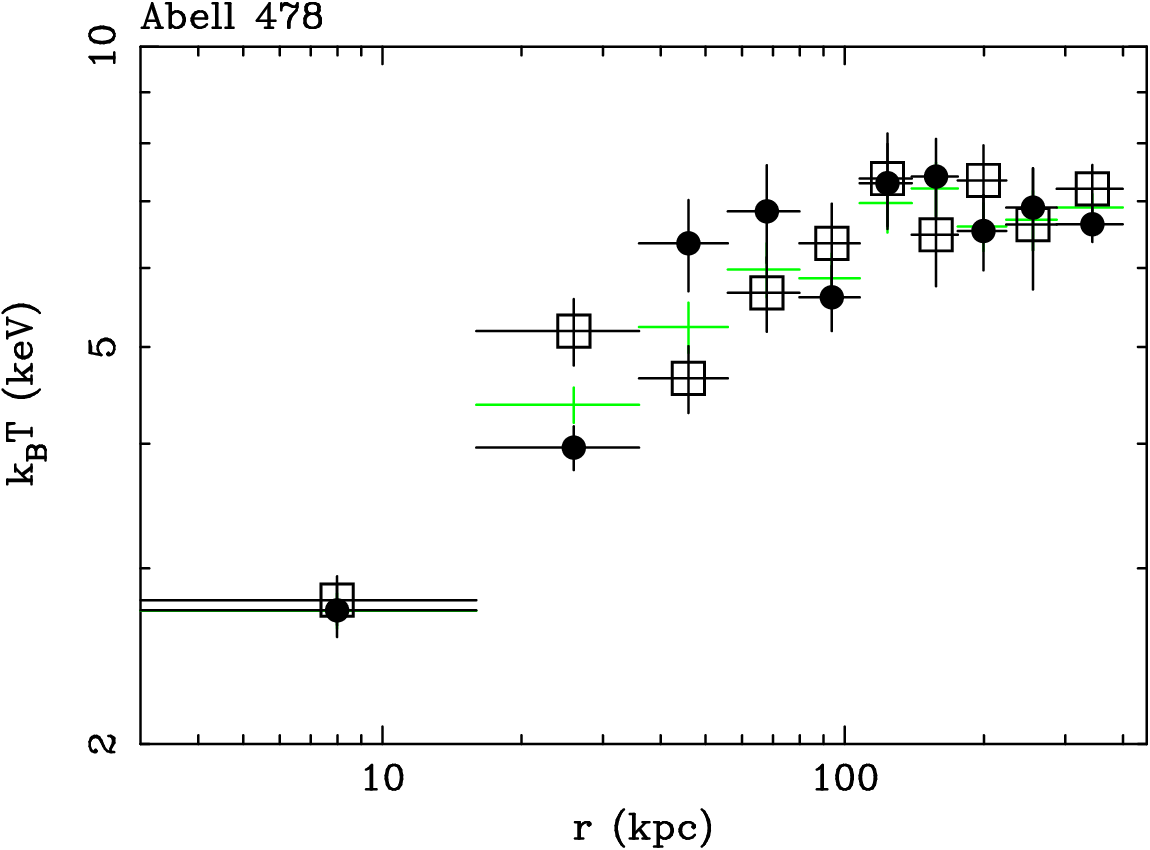}
\caption{Plots showing the temperature distribution extracted from the
  northern (filled circles) and southern (open squares) sectors of the
  cluster. Also shown is the temperature obtained from the whole
  annulus at each radius (green).}
\label{fig:tempSym}
\end{figure*}

Fig~\ref{fig:tempSym} shows the temperature distributions obtained for
each of the clusters showing a flat inner density profile, together
with two clusters which are consistent with the CDM mass model. 

The temperature distribution of 2A~0335$+$096 is clearly asymmetric
within the central $\sim$50 kpc. This is unsurprising given that
\citet{mazzottaetal2003} have shown the central region contains many
blobs of gas at different projected temperatures. The asymmetry does
not, therefore, necessarily mean that the gas is not in hydrostatic
equilibrium, but does suggest that the temperature distribution
obtained from the spectral analysis may not be that of the ambient
gas. In this case, the central mass measurements obtained may be
incorrect. We note that \citet{mazzottaetal2003} find a significant
asymmetry in the temperature distribution at radii $> 100$ kpc. We
also find that the southern half of the cluster is hotter than the
northern half in the outer region, but the effect is not so
pronounced. There are several differences between the analysis carried
out here and the one by \citet{mazzottaetal2003} which may give rise
to a discrepancy in the measured temperature distributions.  Firstly,
\citet{mazzottaetal2003} extract spectra in elliptical rather than
circular annuli and are therefore probing different regions of the
cluster. If the temperature distribution is not symmetrical then this
may lead to significant differences in the temperatures measured. In
addition, \citet{mazzottaetal2003} plot the projected temperatures and
allow the absorption column density to vary between shells.

For the rest of the sample there is little evidence for a correlation
between the inner slope and the degree of symmetry in the temperature
distribution.

\subsection{Observed gas mass fraction profiles}
\begin{figure}
 \centering
\includegraphics[angle=-90,width=0.95\columnwidth]{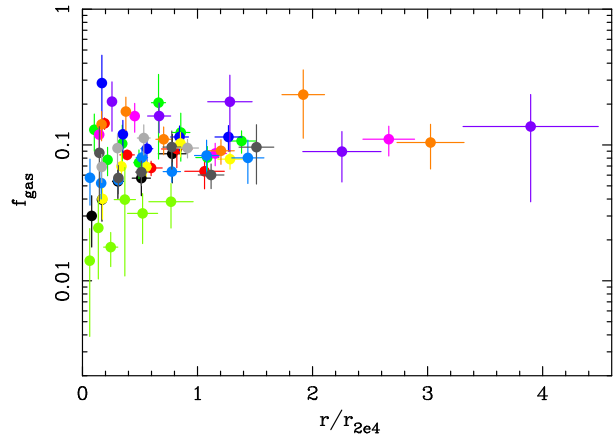}
\caption{Gas mass fraction profiles for the sample with the $x$-axis
  scaled by $r_{2 \times 10^{4}}$ (see Table~\ref{tab:nfw}). The
  colour-coding is the same as used in Fig~\ref{fig:alpha}.}
\label{fig:fgas}
\end{figure}

The gas mass fraction profiles are shown in Fig.~\ref{fig:mass}. The
profiles were calculated using the relation \(f_{\rm gas}(<r)=M_{\rm
  gas}(<r)/M_{\rm tot}(<r)\), where $M_{\rm tot}$ is the total
model-independent mass. The profiles are plotted for the whole sample
in Fig~\ref{fig:fgas}, with the $x$-axis scaled by $r_{2 \times
  10^{4}}$ (see Section 7.4).  The profiles show a variety of
behaviour within $r \sim 0.02r_{\rm vir}$, with an apparent flattening
at $f_{\rm gas} \sim0.1$ outside this region, similar to the results
of \citet{allenetal2004}.

\subsubsection{Inner region}
\begin{figure}
\centering
\vspace{1.0mm}
\includegraphics[angle=-90,width=0.95\columnwidth]{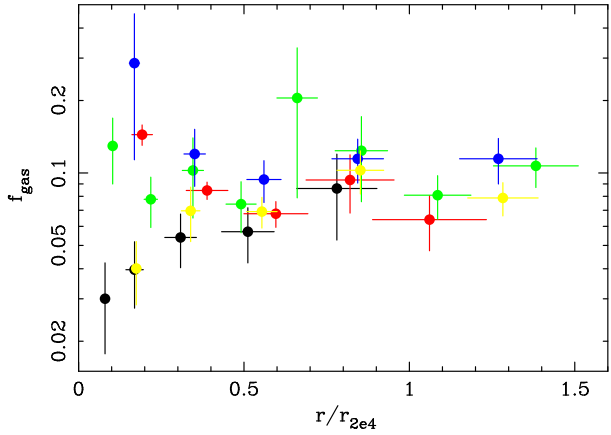}
\vspace{1.0mm}
\caption{Gas mass fraction profiles for Abell~3112 (black),
  2A~0335$+$096 (red), Abell~478 (green), PKS~0745$-$191 (blue) and
  Abell~1795 (yellow).}
\label{fig:fgasinner}
\end{figure}

A high value of $f_{\rm gas}$, relative to the values outside $r \sim
0.02r_{\rm vir}$, is seen in the central bin of 2A~0335$+$096, Abell
478, PKS~0745$-$191 and Sersic~159$-$03.  These are the only objects
in the sample for which the best-fit inner slope, $\alpha$, to the
mass profile is less than 1.0 (see Fig~\ref{fig:alpha}), suggesting a
link between a high \emph{measured} central gas mass fraction and an
\emph{apparently} shallow inner mass density profile.  The central
rise in $f_{\rm gas}$ is clearly seen in Fig.~\ref{fig:fgasinner}
where we plot scaled gas mass fraction profiles for the three clusters
where the $f_{\rm gas}$ peak is most obvious (2A~0335$+$096, Abell~478
and PKS~0745$-$191), together with two low redshift clusters
(Abell~3112 and Abell~1795), whose best-fit mass profiles are
consistent with $\alpha=1.0$, for comparison.  The latter two objects
show an increase in $f_{\rm gas}$ from the centre outwards, as
expected from CDM simulations \citep{ekeetal1998}.  With the exception
of Abell~3581, whose profile drops towards the centre, the remaining
objects in the sample have relatively flat $f_{\rm gas}$ profiles in
the inner region.

\citet{katayama&hayashida2004} find a negative correlation between
$\alpha$ and the value of $f_{\rm gas}$ near the centre of the
cluster.  They interpret this as a tendency for clusters with gas-rich
cores to have flat central density profiles. It seems just as likely
that the high gas mass fraction measured in the central parts of these
objects results from an underestimation of the total mass\footnote{If
  the true gas mass fraction profiles of the objects studied here with
  best-fit $\alpha<1.0$ are either flat or drop into the centre then
  this implies that the total mass in the inner region has been
  underestimated by at least a factor of 2. The total mass profiles
  would then be consistent with $\alpha \sim1.0$.}, supporting the
presence of non-thermal pressure in the core or suggesting incorrect
spectral modelling of the data (note that the spectral fit to the
emission from the centre of 2A~0335$+$096 is poor).

\subsection{NFW model fits}
The majority of the clusters in the sample are consistent with the NFW
mass model i.e. $\alpha = 1.0$ (see Fig.~\ref{fig:alpha}). Here we
find the best-fit scale radius and concentration for each cluster
assuming an NFW profile. (For clusters where the NFW model is excluded
at the 1 sigma level, we remove the inner 30 kpc from the fit). It is
important to stress that attempting to measure $r_{\rm s}$ is pushing
the data to its limits since the best-fit scale radius in general lies
beyond the radius at which mass measurements can be made (see
Fig.~\ref{fig:mass}). The values obtained for $r_{\rm s}$ are
therefore strongly dependent on the outermost datapoint in the mass
profile. It also makes the value difficult to constrain and, while
some of the clusters show evidence for a turn-over in the mass
profile, more precise determination of the scale radius would require
mass measurements out to a larger fraction of the virial radius.

We compute the concentration at three overdensities: $\Delta =
2\times10^{4}$, 2500 and 200. For $\Delta = 2\times10^{4}$ the value
obtained for $r_{\Delta}$ is independent of the accuracy of $r_{\rm
  s}$ since in each case $r_{\Delta}$ lies within the radius at which
mass measurements have been obtained.  However, for $\Delta = 200$ the
value obtained for $r_{200}$ is highly dependent on the accuracy of
$r_{\rm s}$ (since it involves extrapolating the NFW model using
$r_{\rm s}$). The values for $r_{200}$ are shown for interest, but are
not used for any subsequent analysis. The concentration is also
computed at $\Delta = 2500$. This allows comparison of our results for
PKS~0745$-$191, Abell~1835, RXJ~1347.5$-$1145 and MS~2137.3$-$2352
with those found by \citet{allenetaltuni2001}. (Note that for the more
massive clusters $r_{2500}$ lies within the radius at which mass
measurements have been obtained). The data used here for
PKS~0745$-$191, Abell~1835 and RXJ~1347.5$-$1145 are the same data
used by \citet{allenetaltuni2001}. The values of $r_{2500}$ are
consistent between the two studies within the 1 sigma uncertainties.
However, \citet{allenetaltuni2001} generally find a higher
concentration (and therefore lower scale radius) than the results
found here. The difference in values is not surprising given the
discussion above concerning the difficulty of measuring $r_{\rm s}$.

\begin{figure}
\centering
\includegraphics[angle=-90,width=0.95\columnwidth]{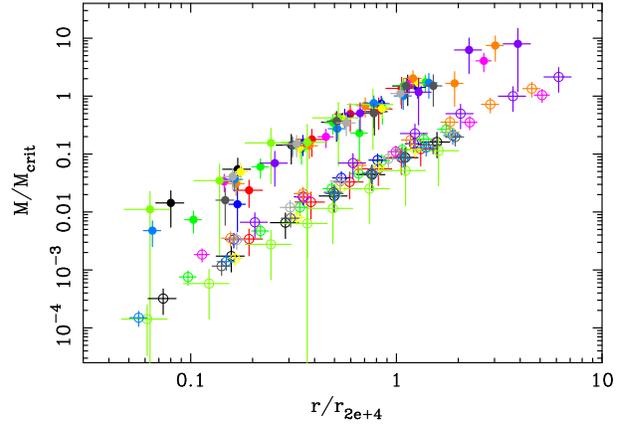}
\caption{Scaled mass profiles: total (filled circles) and gas (open
  circles) mass profiles. The colour-coding is the same as used in
  Fig~\ref{fig:alpha}.}
\label{fig:scaled}
\end{figure}

The best-fit parameters are shown in Table~\ref{tab:nfw}. The mass
enclosed within a density contrast of 2$\times$10$^{4}$ is computed
using Equation~\ref{eqn:rdelta}. The cluster mass profiles, with the
$x$-axis scaled by $r_{\Delta}$, are shown in Fig.~\ref{fig:scaled}.
The $y$-axis for each mass profile is scaled by $M_{\rm crit}=(4/3)\pi
r_{\Delta}^{3}\rho_{\rm crit}\Delta$. Each mass profile should
therefore coincide if the concentration varies little from halo to
halo\footnote{Substituting $r^{\prime}=r/r_{\Delta}$ into
  Equation~\ref{eqn:darkmass} gives \(M(r^{\prime})/M_{\rm
    crit}=\left[\mathrm{ln}(1+c_{\Delta}r^{\prime})-c_{\Delta}r^{\prime}/(1+c_{\Delta}r^{\prime})\right]/\left[\mathrm{ln}(1+c_{\Delta})-c_{\Delta}/(1+c_{\Delta
    })\right]\).}.

\begin{figure}
\centering
\includegraphics[angle=-90,width=0.95\columnwidth]{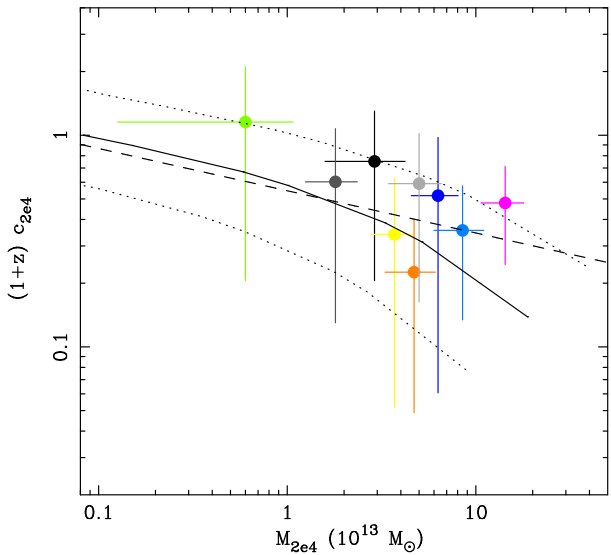}
\caption{Plot showing the dependence of concentration on halo mass.
  The theoretical mass--concentration relation predicted by
  \citet{bullocketal2001} (solid curve) and one-sigma confidence
  limits (dotted curves). Also shown is the best-fit power-law model
  to the data (dashed line). The colour-coding is the same as used in
  Fig~\ref{fig:alpha}.}
\label{fig:massConcRel}
\end{figure}

$N$-body simulations predict a weak dependence of halo concentration
on mass, with $c_{\rm vir} \propto M_{\rm vir}^{-a}$ and $a \sim 0.1$
for all halos at a given redshift. In Fig.~\ref{fig:massConcRel} we
plot $c_{2\times10^4}$ against $M_{2\times10^4}$. (Note that we do not
plot $c_{200}$ against $M_{200}$ here since the calculation of
$M_{200}$ depends on $r_{200}$ which is itself an uncertain number).
The concentration is scaled to a redshift of zero using the
concentration--redshift dependence found in $N$-body simulations by
\citet{bullocketal2001}. The relation predicts that for haloes of the
same mass the concentration is proportional to $(1+z)^{-1}$. We show
the best-fit power-law model to the data along with the theoretical
prediction of \citet{bullocketal2001}\footnote{The model data are
  found using the Fortran codes provided on J.~Bullock's webpage. The
  theoretical $c_{\rm vir}$--$M_{\rm vir}$ relation is scaled to $c_{2
    \times 10^{4}}$--$M_{2 \times 10^{4}}$ using the definition $M_{2
    \times 10^{4}} = M_{\rm vir} \times f(c_{2 \times
    10^{4}})/f(c_{\rm vir})$, where $f(c)=\mathrm{ln}(1+c)-c/(1+c)$,
  $c_{2 \times 10^{4}}=c_{\rm vir} \times (\Delta_{\rm{vir}} M_{2
    \times 10^{4}}/ \Delta_{2 \times 10^{4}} M_{\Delta_{\rm
      vir}})^{1/3}$ and $\Delta_{\rm vir} = 104$.}.  The theoretical
model is applicable out to halo masses of $M_{\rm vir} \lesssim
10^{16} M_{\odot}$, giving rise to a cut-off in the predicted
$c_{2\times10^4}$--$M_{2\times10^4}$ relation at $M_{2\times10^4} \sim
10^{14} M_{\odot}$. The best-fit power-law index is $a \sim 0.2$. The
data appear to agree well with the theoretical model of
\citet{bullocketal2001}.

\begin{table*}
\centering
\begin{tabular}{lcccccccc}
\hline
\\
Cluster & $r_{\rm s}$ & $c_{200}$ & $r_{200}$ & $c_{2\times10^{4}}$ &
$r_{2\times10^{4}}$ & $M_{2\times10^{4}}$&  $r_{2500}$&  $\chi^{2}$(dof)\\
& (Mpc) & & (Mpc) & & (Mpc) &  ($10^{13}$ M$_{\odot}$) & (Mpc)&\\
\hline
Abell~3112 & 0.19$^{+0.39}_{-0.09}$ & 7.06$^{+3.62}_{-3.23}$ &
1.32$^{+0.88}_{-0.35}$ & 0.70$^{+0.57}_{-0.44}$  &
0.13$^{+0.02}_{-0.02}$  & 2.9$^{+1.5}_{-1.1}$  & 0.42$^{+0.19}_{-0.09}$  & 1.3(4
)  \\
\\
2A~0335$+$096 & 0.13$^{+5.38}_{-0.10}$  & 8.18$^{+18.83}_{-7.20}$  &
1.05$^{+4.42}_{-0.39}$  &
0.88$^{+3.22}_{-0.85}$ & 0.11$^{+0.02}_{-0.01}$ & 1.7$^{+1.0}_{-0.5}$
& 0.34$^{+0.56}_{-0.10}$  & 1.7(1)\\
\\
Abell~478 & 1.03$^{+\to\infty}_{-0.63}$  &  2.88$^{+2.02}_{-\to 2.88}$  &
2.97$^{+14.8}_{-1.02}$  & 0.16$^{+0.24}_{-\to 0.16}$   &
0.16$^{+0.02}_{-0.01}$&  5.7$^{+1.7}_{-1.5}$  & 0.76$^{+0.67}_{-0.19}$  & 5.6(5)
\\
\\
PKS~0745$-$191 & 0.36$^{+0.98}_{-0.20}$ & 5.46$^{+3.22}_{-2.88}$  &
1.96$^{+1.51}_{-0.50}$  & 0.47$^{+0.48}_{-0.34}$
& 0.17$^{+0.02}_{-0.01}$  & 6.3$^{+1.9}_{-1.6}$ &0.59$^{+0.27}_{-0.11}$  & 0.6(2
)\\
\\
RXJ~1347.5$-$1145 & 0.60$^{+0.46}_{-0.22}$  & 4.37$^{+1.39}_{-1.24}$  &
2.64$^{+0.70}_{-0.48}$ &
0.33$^{+0.18}_{-0.14}$  & 0.20$^{+0.02}_{-0.02}$  &
14.3$^{+3.6}_{-3.7}$ & 0.76$^{+0.13}_{-0.10}$  & 3.0(2)\\
\\
Abell~1795 & 0.45$^{+1.59}_{-0.23}$ & 4.28$^{+2.23}_{-2.41}$ &
1.94$^{+1.86}_{-0.51}$  &
0.32$^{+0.30}_{-0.24}$  & 0.14$^{+0.01}_{-0.01}$  &
3.7$^{+1.0}_{-0.8}$ &0.55$^{+0.28}_{-0.11}$  & 3.0(3)\\
\\
Abell~1835 & 0.80$^{+1.59}_{-0.38}$  & 3.13$^{+1.37}_{-1.44}$ &
2.50$^{+1.55}_{-0.64}$ &
0.18$^{+0.16}_{-0.12}$  &  0.15$^{+0.01}_{-0.01}$ &
4.7$^{+1.4}_{-1.4}$ & 0.65$^{+0.20}_{-0.12}$  &6.6(4)\\
\\
Abell~3581 & 0.07$^{+0.23}_{-0.04}$  & 9.81$^{+6.30}_{-5.40}$  &
0.69$^{+0.65}_{-0.20}$ & 1.13$^{+1.05}_{-0.79}$
& 0.08$^{+0.02}_{-0.01}$& 0.6$^{+0.6}_{-0.3}$ & 0.23$^{+0.16}_{-0.06}$ &  1.3(4)
\\
\\
Abell~2029 & 0.58$^{+0.94}_{-0.25}$  & 4.38$^{+1.64}_{-1.76}$ &
2.52$^{+1.45}_{-0.59}$ &
0.33$^{+0.22}_{-0.19}$  & 0.19$^{+0.02}_{-0.02}$  &
8.5$^{+2.7}_{-2.3}$& 0.72$^{+0.26}_{-0.13}$  & 2.9(5)\\
\\
RXJ~1532.9$+$3021 & 0.81$^{+13.07}_{-0.53}$  & 2.77$^{+2.28}_{-2.28}$ &
2.24$^{+4.72}_{-0.85}$ &
0.15$^{+0.27}_{-0.14}$  & 0.12$^{+0.02}_{-0.02}$ & 2.9$^{+1.3}_{-1.3}$
& 0.56$^{+0.31}_{-0.15}$ & 1.3(3)\\
\\
MS~2137.3$-$2353 & 0.34$^{+0.74}_{-0.16}$ & 5.28$^{+2.41}_{-2.52}$ &
1.74$^{+1.22}_{-0.42}$ & 0.45$^{+0.35}_{-0.30}$ &
0.15$^{+0.02}_{-0.01}$ & 5.0$^{+1.7}_{-1.4}$ & 0.52$^{+0.22}_{-0.10}$  & 0.4(2)
\\
\\
Sersic~159$-$03 & 0.20$^{+0.37}_{-0.10}$  & 6.16$^{+3.42}_{-2.79}$ &
1.22$^{+0.70}_{-0.30}$ & 0.57$^{+0.52}_{-0.36}$  &
0.11$^{+0.01}_{-0.01}$ & 1.8$^{+0.6}_{-0.5}$& 0.38$^{+0.14}_{-0.07}$& 2.5(3)\\
\hline
\end{tabular}
\caption{Best-fit NFW model parameters for $\Delta=200$ and
  $\Delta=2\times 10^{4}$. Also shown are the scaled radii enclosing a
  density contrast of $\Delta=2500$.}
\label{tab:nfw}
\end{table*}

\section{Summary}
The shape of the mass distribution in virialized halos provides a
sensitive test of the CDM paradigm. In this paper we have analysed the
mass profiles of a sample of galaxy clusters. The main results are as
follows:
 
\begin{itemize}
\item The mass distribution in the majority of the objects studied are
  consistent with the NFW model \citep{nfw1995}. Four objects in the
  sample (2A~0335$+$096, Abell~478, PKS~0745$-$191 and
  Sersic~159$-$03) exhibit a flatter core.
 
\item The gas mass fraction measured in the central region of the
  clusters with flat core density profiles is high, indicating that
  the total mass may be underestimated within the central $\sim 30
  \kpc$ of these objects. If this is the case, then either a
  significant non-thermal pressure component exists in the core, or
  the X-ray emission has been incorrectly modelled. For 2A~0335$+$096
  the central temperature distribution is asymmetric, suggesting that
  the temperature and density obtained may not accurately represent
  the ambient gas properties in the core for this object.

\item CDM simulations predict that the mass distribution in virialized
  halos differs from a power-law, turning-over at a characteristic
  radius, \(r_{\rm s}=r_{\rm vir}/c\).
\begin{itemize}
\item With $\alpha$ a free parameter in the fits, a power-law density
  profile is ruled out at the 1 sigma level in Abell~1835 and Abell
  2029. (A power-law is also ruled out at the 1 sigma level in 2A
  0335+096, Abell~478, PKS~0745$-$191 and Sersic~159$-$03, although
  the inner slope in these objects is inconsistent with CDM
  predictions).  Otherwise, we find no strong evidence for the
  expected turn-over in the mass profile.  For three objects in the
  sample (Abell~3112, Abell~1795 and MS~2137.3$-$2353) the best-fit
  model is a power-law.
\item With $\alpha$ fixed at 1.0 the best-fit model is inconsistent
  with a power-law model for all objects except Abell~478. 
\end{itemize}

\item The mass--concentration relation at a density contrast of
  $2\times10^{4}$ is consistent with the theoretical prediction from
  $N$-body simulations by \citet{bullocketal2001}.
\end{itemize}

\section{Discussion}
In the CDM scenario, simulations predict a universal form for the
distribution of dark matter in virialized halos, independent of both
mass and cosmological parameters \citep{nfw1995,nfw1996,nfw1997}. The
universal density profile has a cuspy core, $\rho \propto
r^{-\alpha}$, rolling over to $\rho \propto r^{-3}$ at a
characteristic radius known as the scale radius.  The majority of the
mass profiles obtained for the cluster sample in this paper are
consistent with the NFW model; only four objects exhibit shallower
density cores (2A~0335$+$096, Abell~478, PKS~0745$-$191 and
Sersic~159$-$03).  The discrepancy between the mass distribution
measured in these systems and the NFW profile need not indicate any
problem with the CDM paradigm, but may instead result from an
underestimation of the core mass. The high gas mass fraction measured
in the central region of all four systems with flat inner density
profiles certainly suggests that this may be the case. Spuriously low
mass measurements could be caused by the presence of non-thermal
pressure support in the inner regions of these objects and/or
incorrect spectral modelling of the emission from clusters with
significant core substructure. Another possibility concerns the
importance of non-spherical effects on the measured mass distribution.
\citet{hayashietaltriaxial2004} have recently shown that rotation
curves obtained from galaxies with triaxial halos may, in some cases,
infer an erroneously flat core to the mass profile if spherical
symmetry is assumed. The effects of non-axial symmetry on X-ray
measurements of cluster masses have been investigated by
\citet{piffarettietal2003}. The authors show that the core mass will
be underestimated if the halo is compressed along the line-of-sight.
It is interesting that both Abell~478 and PKS~0745$-$191 both appear
highly elliptical.  Analytic solutions to the hydrostatic equation for
gas residing in a triaxial dark matter potential have been computed
for isothermal and polytropic temperature distributions by
\citet{lee&suto2003}. These formula may in future work be applied to
clusters to compute the mass distribution.

Determining whether or not the flat cores measured in some clusters
are real is of great importance to cosmological and cluster studies;
if they \emph{are} real, then either our understanding of dark matter
is incorrect, or non-gravitational processes, such as dynamical
friction acting on cluster galaxies moving through the core
\citep{elzantetal2004}, modify the mass distribution.

The work carried out here has put some constraints on the mass
distribution in a sample of galaxy clusters. Tighter constraints will
be possible by imaging the clusters out to larger radii or, for the
case of high redshift objects, by taking longer exposures, so that the
radius at which the profile `turns-over' may be robustly detected.

In future work it will be important to investigate effects which might
reconcile the observed flat density profiles found in this study with
CDM model predictions. One possible cause for the discrepancy is
non-thermal pressure support in the core. \citet{sandersetal2005} find
inverse Compton emission from relativistic electrons which can
contribute significant pressure near the centre of the Perseus
cluster. Careful comparison of X-ray and lensing masses will also be
important.

\section*{Acknowledgements}
LMV is grateful to Roderick Johnstone and Steve Allen for their help
and advice. We thank the referee for several comments which led to a
significantly improved manuscript. ACF and LMV acknowledge support
from The Royal Society and PPARC, respectively.

\bibliographystyle{mnras}                       
\bibliography{mn-jour,references_c1,references_c2,references_c3,references_c4,references_c5}

\begin{thebibliography}{}

\bibitem[\protect\citeauthoryear{{Allen}}{{Allen}}{1998}]{allen1998}
{Allen} S.~W., 1998, \mnras, 296, 392

\bibitem[\protect\citeauthoryear{{Allen} et~al.}{{Allen}
  et~al.}{2004}]{allenetal2004}
{Allen} S.~W., {Schmidt} R.~W., {Ebeling} H., {Fabian} A.~C.,  {van Speybroeck}
  L., 2004, \mnras, 353, 457

\bibitem[\protect\citeauthoryear{{Allen}, {Schmidt}, \& {Fabian}}{{Allen}
  et~al.}{2001}]{allenetaltuni2001}
{Allen} S.~W., {Schmidt} R.~W.,  {Fabian} A.~C., 2001, \mnras, 328, L37

\bibitem[\protect\citeauthoryear{{Anders} \& {Grevesse}}{{Anders} \&
  {Grevesse}}{1989}]{anders&grev1989}
{Anders} E.,  {Grevesse} N., 1989, \gca, 53, 197

\bibitem[\protect\citeauthoryear{{Arabadjis}, {Bautz}, \&
  {Arabadjis}}{{Arabadjis} et~al.}{2004}]{aba2004}
{Arabadjis} J.~S., {Bautz} M.~W.,  {Arabadjis} G., 2004, \apj, 617, 303

\bibitem[\protect\citeauthoryear{{Arnaud}}{{Arnaud}}{1996}]{arnaud96}
{Arnaud} K., 1996, in {Astronomical Society of the Pacific conference series},
  Vol. 101, {Jacoby} G.~H.,  {Barnes} J., ed, {Astronomical Data Analysis
  Software and Systems}, p.~17

\bibitem[\protect\citeauthoryear{{B{\^ i}rzan} et~al.}{{B{\^ i}rzan}
  et~al.}{2004}]{birzanetal2004}
{B{\^ i}rzan} L., {Rafferty} D.~A., {McNamara} B.~R., {Wise} M.~W.,  {Nulsen}
  P.~E.~J., 2004, \apj, 607, 800

\bibitem[\protect\citeauthoryear{{Balucinska-Church} \&
  {McCammon}}{{Balucinska-Church} \& {McCammon}}{1992}]{bal&mc1992}
{Balucinska-Church} M.,  {McCammon} D., 1992, \apj, 400, 699

\bibitem[\protect\citeauthoryear{{Binney} \& {Tremaine}}{{Binney} \&
  {Tremaine}}{1987}]{binney&tremaine1987}
{Binney} J.,  {Tremaine} S., 1987, {Galactic dynamics}.
\newblock Princeton University Press, Princeton, NJ

\bibitem[\protect\citeauthoryear{{Bullock} et~al.}{{Bullock}
  et~al.}{2001}]{bullocketal2001}
{Bullock} J.~S., {Kolatt} T.~S., {Sigad} Y., {Somerville} R.~S., {Kravtsov}
  A.~V., {Klypin} A.~A., {Primack} J.~R.,  {Dekel} A., 2001, \mnras, 321, 559

\bibitem[\protect\citeauthoryear{{Cole} \& {Lacey}}{{Cole} \&
  {Lacey}}{1996}]{cole&lacey1996}
{Cole} S.,  {Lacey} C., 1996, \mnras, 281, 716

\bibitem[\protect\citeauthoryear{{Crone}, {Evrard}, \& {Richstone}}{{Crone}
  et~al.}{1994}]{croneetal1994}
{Crone} M.~M., {Evrard} A.~E.,  {Richstone} D.~O., 1994, \apj, 434, 402

\bibitem[\protect\citeauthoryear{{Dickey} \& {Lockman}}{{Dickey} \&
  {Lockman}}{1990}]{dick&lock1990}
{Dickey} J.~M.,  {Lockman} F.~J., 1990, \araa, 28, 215

\bibitem[\protect\citeauthoryear{{Diemand} et~al.}{{Diemand}
  et~al.}{2005}]{diemandetal2005}
{Diemand} J., {Zemp} M., {Moore} B., {Stadel} J.,  {Carollo} M., 2005, ArXiv
  Astrophysics e-prints

\bibitem[\protect\citeauthoryear{{Dubinski} \& {Carlberg}}{{Dubinski} \&
  {Carlberg}}{1991}]{dubinski&carlberg1991}
{Dubinski} J.,  {Carlberg} R.~G., 1991, \apj, 378, 496

\bibitem[\protect\citeauthoryear{{Eke}, {Cole}, \& {Frenk}}{{Eke}
  et~al.}{1996}]{ekeetal1996}
{Eke} V.~R., {Cole} S.,  {Frenk} C.~S., 1996, \mnras, 282, 263

\bibitem[\protect\citeauthoryear{{Eke}, {Navarro}, \& {Frenk}}{{Eke}
  et~al.}{1998}]{ekeetal1998}
{Eke} V.~R., {Navarro} J.~F.,  {Frenk} C.~S., 1998, \apj, 503, 569

\bibitem[\protect\citeauthoryear{{El-Zant} et~al.}{{El-Zant}
  et~al.}{2004}]{elzantetal2004}
{El-Zant} A.~A., {Hoffman} Y., {Primack} J., {Combes} F.,  {Shlosman} I., 2004,
  \apjl, 607, L75

\bibitem[\protect\citeauthoryear{{Fillmore} \& {Goldreich}}{{Fillmore} \&
  {Goldreich}}{1984}]{fillmore&goldreich1984}
{Fillmore} J.~A.,  {Goldreich} P., 1984, \apj, 281, 9

\bibitem[\protect\citeauthoryear{{Frenk} et~al.}{{Frenk}
  et~al.}{1988}]{frenketal1988}
{Frenk} C.~S., {White} S.~D.~M., {Davis} M.,  {Efstathiou} G., 1988, \apj, 327,
  507

\bibitem[\protect\citeauthoryear{{Fukushige} \& {Makino}}{{Fukushige} \&
  {Makino}}{1997}]{fukushige&makino1997}
{Fukushige} T.,  {Makino} J., 1997, \apjl, 477, L9

\bibitem[\protect\citeauthoryear{{Fukushige} \& {Makino}}{{Fukushige} \&
  {Makino}}{2001}]{fukushige&makino2001}
{Fukushige} T.,  {Makino} J., 2001, \apj, 557, 533

\bibitem[\protect\citeauthoryear{{Fukushige} \& {Makino}}{{Fukushige} \&
  {Makino}}{2003}]{fukushige&makino2003}
{Fukushige} T.,  {Makino} J., 2003, \apj, 588, 674

\bibitem[\protect\citeauthoryear{{Ghigna} et~al.}{{Ghigna}
  et~al.}{2000}]{ghignaetal2000}
{Ghigna} S., {Moore} B., {Governato} F., {Lake} G., {Quinn} T.,  {Stadel} J.,
  2000, \apj, 544, 616

\bibitem[\protect\citeauthoryear{{Gunn} \& {Gott}}{{Gunn} \&
  {Gott}}{1972}]{gunn&gott1972}
{Gunn} J.~E.,  {Gott} J.~R.~I., 1972, \apj, 176, 1

\bibitem[\protect\citeauthoryear{{Hayashi} et~al.}{{Hayashi}
  et~al.}{2004}]{hayashietaltriaxial2004}
{Hayashi} E. et~al., 2004, ApJ, submitted (astro-ph/0408132)

\bibitem[\protect\citeauthoryear{{Hayashi} et~al.}{{Hayashi}
  et~al.}{2003}]{hayashietal2003}
{Hayashi} E. et~al., 2003, MNRAS, submitted (astro-ph/0310576)

\bibitem[\protect\citeauthoryear{{Hoffman} \& {Shaham}}{{Hoffman} \&
  {Shaham}}{1985}]{hoffman&shaham1985}
{Hoffman} Y.,  {Shaham} J., 1985, \apj, 297, 16

\bibitem[\protect\citeauthoryear{{Jing} \& {Suto}}{{Jing} \&
  {Suto}}{2000}]{jing&suto2000}
{Jing} Y.~P.,  {Suto} Y., 2000, \apjl, 529, L69

\bibitem[\protect\citeauthoryear{{Johnstone} et~al.}{{Johnstone}
  et~al.}{2004}]{johnstoneetal2004}
{Johnstone} R.~M., {Fabian} A.~C., {Morris} R.~G.,  {Taylor} G.~B., 2004,
  MNRAS, accepted (astro-ph/0410154)

\bibitem[\protect\citeauthoryear{{Katayama} \& {Hayashida}}{{Katayama} \&
  {Hayashida}}{2004}]{katayama&hayashida2004}
{Katayama} H.,  {Hayashida} K., 2004, Advances in Space Research, 34, 2519

\bibitem[\protect\citeauthoryear{{Klypin} et~al.}{{Klypin}
  et~al.}{2001}]{klypinetal2001}
{Klypin} A., {Kravtsov} A.~V., {Bullock} J.~S.,  {Primack} J.~R., 2001, \apj,
  554, 903

\bibitem[\protect\citeauthoryear{{Kravtsov}, {Klypin}, \&
  {Khokhlov}}{{Kravtsov} et~al.}{1997}]{kravtsovetal1997}
{Kravtsov} A.~V., {Klypin} A.~A.,  {Khokhlov} A.~M., 1997, \apjs, 111, 73

\bibitem[\protect\citeauthoryear{{Lacey} \& {Cole}}{{Lacey} \&
  {Cole}}{1993}]{lacey&cole1993}
{Lacey} C.,  {Cole} S., 1993, \mnras, 262, 627

\bibitem[\protect\citeauthoryear{{Lee} \& {Suto}}{{Lee} \&
  {Suto}}{2003}]{lee&suto2003}
{Lee} J.,  {Suto} Y., 2003, \apj, 585, 151

\bibitem[\protect\citeauthoryear{{Liedahl}, {Osterheld}, \&
  {Goldstein}}{{Liedahl} et~al.}{1995}]{lied&ost&gold1995}
{Liedahl} D.~A., {Osterheld} A.~L.,  {Goldstein} W.~H., 1995, \apjl, 438, L115

\bibitem[\protect\citeauthoryear{{Mazzotta}, {Edge}, \&
  {Markevitch}}{{Mazzotta} et~al.}{2003}]{mazzottaetal2003}
{Mazzotta} P., {Edge} A.~C.,  {Markevitch} M., 2003, \apj, 596, 190

\bibitem[\protect\citeauthoryear{{Mewe}, {Gronenschild}, \& {van den
  Oord}}{{Mewe} et~al.}{1985}]{mew&gron&van1985}
{Mewe} R., {Gronenschild} E.~H.~B.~M.,  {van den Oord} G.~H.~J., 1985, \aaps,
  62, 197

\bibitem[\protect\citeauthoryear{{Moore} et~al.}{{Moore}
  et~al.}{1999}]{mooreetal1999}
{Moore} B., {Quinn} T., {Governato} F., {Stadel} J.,  {Lake} G., 1999, \mnras,
  310, 1147

\bibitem[\protect\citeauthoryear{{Navarro}, {Frenk}, \& {White}}{{Navarro}
  et~al.}{1995}]{nfw1995}
{Navarro} J.~F., {Frenk} C.~S.,  {White} S.~D.~M., 1995, \mnras, 275, 720

\bibitem[\protect\citeauthoryear{{Navarro}, {Frenk}, \& {White}}{{Navarro}
  et~al.}{1996}]{nfw1996}
{Navarro} J.~F., {Frenk} C.~S.,  {White} S.~D.~M., 1996, \apj, 462, 563

\bibitem[\protect\citeauthoryear{{Navarro}, {Frenk}, \& {White}}{{Navarro}
  et~al.}{1997}]{nfw1997}
{Navarro} J.~F., {Frenk} C.~S.,  {White} S.~D.~M., 1997, \apj, 490, 493

\bibitem[\protect\citeauthoryear{{Navarro} et~al.}{{Navarro}
  et~al.}{2004}]{navarroetal2004}
{Navarro} J.~F. et~al., 2004, \mnras, 349, 1039

\bibitem[\protect\citeauthoryear{{Piffaretti}, {Jetzer}, \&
  {Schindler}}{{Piffaretti} et~al.}{2003}]{piffarettietal2003}
{Piffaretti} R., {Jetzer} P.,  {Schindler} S., 2003, \aap, 398, 41

\bibitem[\protect\citeauthoryear{{Pointecouteau}, {Arnaud}, \&
  {Pratt}}{{Pointecouteau} et~al.}{2005}]{pap2005}
{Pointecouteau} E., {Arnaud} M.,  {Pratt} G.~W., 2005, \aap, 435, 1

\bibitem[\protect\citeauthoryear{{Power} et~al.}{{Power}
  et~al.}{2003}]{poweretal2003}
{Power} C., {Navarro} J.~F., {Jenkins} A., {Frenk} C.~S., {White} S.~D.~M.,
  {Springel} V., {Stadel} J.,  {Quinn} T., 2003, \mnras, 338, 14

\bibitem[\protect\citeauthoryear{{Sanders}, {Fabian}, \& {Dunn}}{{Sanders}
  et~al.}{2005}]{sandersetal2005}
{Sanders} J.~S., {Fabian} A.~C.,  {Dunn} R.~J.~H., 2005, \mnras, 360, 133

\bibitem[\protect\citeauthoryear{{Syer} \& {White}}{{Syer} \&
  {White}}{1998}]{syer&white1998}
{Syer} D.,  {White} S.~D.~M., 1998, \mnras, 293, 337

\bibitem[\protect\citeauthoryear{{Tormen}, {Bouchet}, \& {White}}{{Tormen}
  et~al.}{1997}]{tormenetal1997}
{Tormen} G., {Bouchet} F.~R.,  {White} S.~D.~M., 1997, \mnras, 286, 865

\bibitem[\protect\citeauthoryear{{Voigt} \& {Fabian}}{{Voigt} \&
  {Fabian}}{2004}]{voigt&fabian2004}
{Voigt} L.~M.,  {Fabian} A.~C., 2004, \mnras, 347, 1130

\bibitem[\protect\citeauthoryear{{Wambsganss}, {Bode}, \&
  {Ostriker}}{{Wambsganss} et~al.}{2005}]{wambetal2005}
{Wambsganss} J., {Bode} P.,  {Ostriker} J.~P., 2005, \apjl, 635, L1

\bibitem[\protect\citeauthoryear{{White} \& {Zaritsky}}{{White} \&
  {Zaritsky}}{1992}]{white&zaritsky1992}
{White} S.~D.~M.,  {Zaritsky} D., 1992, \apj, 394, 1

\end{thebibliography}
\end{document}